\documentclass[12pt,preprint]{aastex}
\shorttitle{[OIII] images of Seyferts}
\shortauthors{Schmitt et al.}
\begin{document}
 
\title{Ultraviolet--to--Far--Infrared Properties of Local Star--Forming
Galaxies\altaffilmark{1,2}}

\author{H. R. Schmitt\altaffilmark{3,4,5}, D. Calzetti\altaffilmark{6}, L.
Armus\altaffilmark{7}, M. Giavalisco\altaffilmark{6},
T. M. Heckman\altaffilmark{6,10}, R. C. Kennicutt Jr.\altaffilmark{8,9},
C. Leitherer\altaffilmark{6}, and G. R. Meurer\altaffilmark{10} }
\altaffiltext{1}{Based on observations made with the NASA/ESA Hubble Space
Telescope, which is operated by the Association of Universities for Research
in Astronomy, Inc., under NASA contract NAS5-26555.}
\altaffiltext{2}{Based on observations obtained with the Apache Point Observatory
3.5-meter telescope, which is owned and operated by the Astrophysical Research
Consortium.}
\altaffiltext{3}{Remote Sensing Division, Code 7210, Naval Research Laboratory,
4555 Overlook Avenue, Washington, DC 20375}
\altaffiltext{4}{Interferometrics, Inc., 13454 Sunrise Valley Drive, Suite 240, Herndon,
VA\,20171}
\altaffiltext{5}{email:hschmitt@ccs.nrl.navy.mil}
\altaffiltext{6}{Space Telescope Science Institute, 3700 San Martin Drive, Baltimore, 
MD21218}
\altaffiltext{7}{Spitzer Science Center, California Institute of Technology,
Mail Stop 220-6, Pasadena, CA 91125}
\altaffiltext{8}{Steward Observatory, University of Arizona, 933 North Cherry
Avenue, Tucson, AZ 85721}
\altaffiltext{9}{Institute of Astronomy, University of Cambridge, Madingley
Road, Cambridge CB3 0HA, UK}
\altaffiltext{10}{Department of Physics and Astronomy,
Johns Hopkins University, Baltimore, MD21218}

\begin{abstract}
We present the results of a multiwavelength study of nearby galaxies,
aimed at understanding the relation between the ultraviolet and
far--infrared emission in star--forming galaxies. The dataset
comprises new ultraviolet (from HST/STIS), ground--based H$\alpha$,
and radio continuum observations, together with archival infrared data
(from IRAS and ISO). The local galaxies are used as benchmarks for
comparison of the infrared--to--ultraviolet properties with two
populations of high--redshift galaxies: the sub--millimeter
star--forming galaxies detected by SCUBA and the ultraviolet--selected
Lyman Break galaxies. In addition, the long wavelength baseline
covered by the present data enables us to compare the star formation
rates (SFRs) derived from the observed ultraviolet, H$\alpha$,
infrared, and radio luminosities, and to gauge the impact of dust opacity
in the local galaxies. We also derive a new calibration for the non-thermal
part of the radio SFR estimator, based on the comparison of 1.4~GHz
measurements with a new estimator of the bolometric luminosity of the star
forming regions. We find that more actively star forming galaxies show higher
dust opacities, in line with previous results. We find that the local
star--forming galaxies have lower F$_{\lambda}$(205~$\mu$m)/F$_{\lambda}$(UV)
ratio, by two--three orders of magnitude than the submillimeter--selected
galaxies, and may have similar or somewhat higher
F$_{\lambda}$(205~$\mu$m)/F$_{\lambda}$(UV) than Lyman Break Galaxies. The
F$_{\lambda}$(205~$\mu$m)/F$_{\lambda}$(UV) ratio of the local galaxy
population may be influenced
by the cool dust emission in the far--infrared heated by non-ionizing
stellar populations, which may be reduced or absent in the LBGs.
\end{abstract}
 

\section{Introduction}

The presence of observational and/or physical links between local
star--forming galaxies and high--redshift ultraviolet--selected and
infrared--selected galaxies is still subject of scrutiny, in view of their
importance for placing the high--redshift populations in the context
of galaxy evolution. One of the outstanding questions is how the
observed UV (restframe $\sim$1600~\AA) and far--infrared (restframe
200--260~$\mu$m) properties of the high--redshift populations relate
to the analogous properties of local star--forming galaxies.

In galaxies, the restframe UV emission traces massive stars and the
recent star formation, modulo the effects of dust opacity. The far
infrared emission at 200--260~$\mu$m traces the Rayleigh--Jeans tail
of the dust emission. Evolved stellar populations unassociated with
the recent star formation may heat the dust to relatively cool
temperatures (T$\lesssim$20~K, e.g. Helou 1986; Lonsdale-Persson \&
Helou 1987; Rowan-Robinson \& Crawford 1989; Rowan-Robinson \&
Efstathiou 1993), which can provide a significant
contribution to the emission at the long wavelengths. The relevance of
investigating F$_{\lambda}$(200--260~$\mu$m) comes from the discovery in recent
years of a significant population of submillimeter--bright sources at
high redshifts observed with SCUBA (Chapman et al. 2003, 2005; Wang,
Cowie, \& Barger 2004; Aretxaga, Hughes, \& Dunlop 2005; Smail et
al. 1997; Barger et al. 1998, 2000; Blain et al. 1999a,b). SCUBA is
most sensitive at 850~$\mu$m, which corresponds to restframe
210--260$\mu$m at redshift z$\sim$2.2--3.

The ultraviolet--selected Lyman Break galaxies at z$\sim$3 (LBGs,
Steidel et al. 1999) resemble local UV--bright starburst galaxies in
many of their UV spectral properties (Meurer, Heckman, \& Calzetti 
1999; Adelberger, \& Steidel 2000).
The LBGs have been argued to be the major
contributor of the global star formation at redshift $\sim$2--4
(Adelberger \& Steidel 2000, Giavalisco 2002; Giavalisco et al. 2004),
although this
has been recently challenged by Chapman et al.(2005). Chapman et
al. (2005) have suggested that the sub--millimeter--detected galaxies
at high redshift (SCUBA sources, median z$\sim$2.2) represent a distinct and
complementary population to the LBGs; the authors also argue that this
population provides a significant contribution to the SFR density of
the Universe in the redshift range z$\sim$2--3.

The sub-mm selected sources have been linked to local Ultra Luminous
Infrared Galaxies (ULIGs, Smail et al. 1997; Blain et al. 1999a,b),
with star formation rates per unit area close to the 'maximum
starburst limit' of Lehnert \& Heckman (1996), Meurer et al. (1997)
and total rates of many
100's M$_{\odot}$~yr$^{-1}$ (Barger et al. 1998; Chapman et al. 2004,
2005; Hughes et al. 1998). Like ULIGs, SCUBA sources tend to be faint
in the restframe ultraviolet (Chapman et al. 2005), but unlike ULIGs,
many SCUBA sources show evidence of extended star formation over many
kpc (Chapman et al. 2004).

Conversely, the z$\sim$3 LBGs are faint, typically undetected, in the
SCUBA waveband, with fluxes around or below the 1~mJy level (Chapman
et al. 2000). As previously suggested (Chapman et al. 2005, and
references therein), the LBGs and the SCUBA sources likely represent
complementary facets of the high--redshift star formation: UV--bright
and 210~$\mu$m--faint the first, and UV--faint and
210--260~$\mu$m--bright the second, perhaps marking a continuum of
properties similar to that in the local Universe between UV--bright
starbursts and ULIGs.

Relating these characteristics to those of local populations has
proceeded so far in a piecemeal fashion with small samples of nearby
galaxies observed simultaneously in the UV and the restframe
$\sim$200~$\mu$m. For this reason we obtained HST/STIS--UV
(1600\AA) observations of a sample of local star--forming galaxies
for which archival ISO observations at $\lambda>$170~$\mu$m
existed. The sample is large enough to offer a unique opportunity to
compare the UV/FIR properties of local galaxies with those of
observations in similar wavebands of high--redshift galaxies.

The availability of an homogeneous set of new ultraviolet, H$\alpha$
and radio data, augmented with IRAS/ISO infrared data, enables us also
to investigate the impact of dust opacity on UV and optical SFR
indicators. There are extensive studies on the subject performed on a
variety of samples of local galaxies (e.g., Sullivan et al. 2001,
Kewley et al. 2002, 2004, Rosa Gonz\'alez et al. 2002, Hopkins et
al. 2003, Bell 2003). Kewley et al. (2002, 2004) uses SFR(IR) as a
benchmark for analyzing SFR(H$\alpha$) and SFR([OII]) for galaxies in
the Nearby Field Galaxy Survey. A similar approach is used by Rosa
Gonz\'alez et al. (2002), while Hopkins et al. (2003) use SFR(1.4~GHz)
as a reference to investigate optical SFR estimators for the galaxies
of the Sloan Digital Sky Survey. A similar study to ours has been
undertaken by Bell (2003), using however a less homogeneous
dataset. A major study on the cross--correlation of multiwavelength
SFR estimators is being undertaken by the SINGS project (the Spitzer
Infrared Nearby Galaxies Survey, Kennicutt et al. 2003), that will
combine data from the B-band through the near-IR, mid-IR and far-IR
(up to 160$\mu$m), all the way to the radio; SINGS will provide a complete,
homogeneous dataset for this type of investigation. The present
study is at the same time complementary to and independent of SINGS,
as there is minimal overlap between the galaxy samples (only three
objects in common). We derive a new calibration for SFR(radio), based
on the comparison of this indicator with a new indicator of the bolometric
luminosity of the star forming regions, which represents a better approximation
of the actual bolometric luminosity than simply using the infrared luminosity. 
These results are compared with other empirically derived calibrations based
on the radio--FIR correlation (Yun et al. 2001, Bell 2003).

This paper is organized in the following way. In Section 2 we present
the data and sample being used. In Section 3 we present the star
formation rate calibrations employed. A
comparison between the different SFRs is given in Section 4, where we
also present an improved radio SFR calibration. Section 5 presents the IR/UV
properties of the sample and compares it to high-redshift galaxies.
A summary of the results is given in Section 6. We assume
H$_o=75$~km~s$^{-1}$~Mpc$^{-1}$ throughout this paper, and, where
necessary, convert data from other papers to this value.

\section{The Data}

\subsection{Nearby Galaxies}

The data and measurements of nearby star forming galaxies used in the
current paper are presented in Paper I (Schmitt et al. 2005).  Our
sample consists of 41 galaxies spanning a wide range in the intrinsic
parameters of luminosity, star formation rate and metallicity. We point
out that these galaxies were culled from the ISO archive, so the sample
may still have some selection effects and may not necessarily represent
typical galaxies, like a volume limited sample.
Measurements were made in the UV (1600\AA) with HST/STIS,
in H$\alpha$ with ground-based facilities and HST archival data, and
in the radio at 8.46~GHz, 4.89~GHz and 1.4~GHz (3.6~cm, 6~cm and
20~cm, respectively), with the VLA and literature data. Archival IRAS
and ISO data were utilized for the infrared; 29 out of 41 galaxies
presented here have ISO data at wavelengths longward of 170~$\mu$m
(Schmitt et al. 2005), thus providing a direct comparison with SCUBA
data for objects at redshift z$\sim$2-3. More details about the data
reductions, measurements and characteristics are given in Paper~I. The
UV data were obtained with an aperture that is about 25\arcsec\ in
side, which is smaller than the typical size of our galaxies. In the
present paper we will only compare UV data with H$\alpha$ and radio
8.46~GHz data measured inside an aperture that matches the UV one.
The infrared data encompass the entire galaxy, and care will be taken
when comparing these measurements to those at other wavelengths, so as
to mitigate the effects of aperture mismatch.

\begin{figure}
\plotone{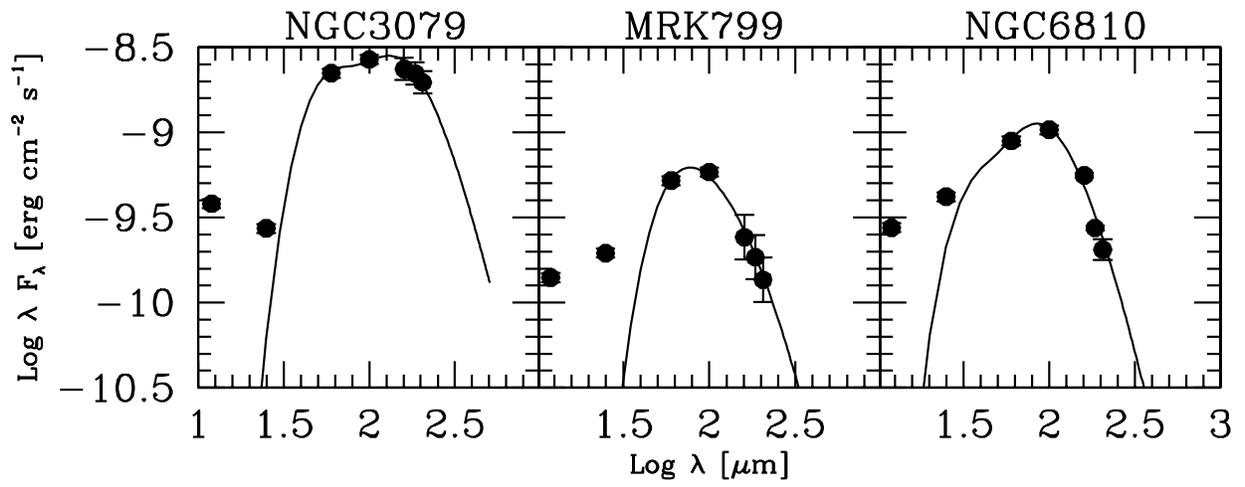}
\caption{Examples of two temperature fit models done to galaxies with
IRAS and ISO measurements.}
\label{fig1}
\end{figure}

In the case of UV, H$\alpha$ and radio data we simply use the
monochromatic luminosities to convert to star formation rates, as
detailed in the next section. In the case of the infrared emission we
integrate the area under the spectral energy distribution between
8$\mu$m and 1~mm to derive star formation rates. This is
accomplished by fitting two temperature models to the data of those
galaxies (see Table~5 in paper I) that had at least one measurement
from ISO longward of 100$\mu$m.  We assumed that the dust emissivity
has index $\epsilon=2$. We present in Figure~\ref{fig1} the example of
three of these fits. The agreement between the fit and observed points
is usually very good. Fitting two temperature models with fixed
emissivity $\epsilon$ allows more flexible SEDs than a single
temperature with variable $\epsilon$, since in the former case two
maxima are one of the possible solutions (e.g. NGC\,3079 Figure~1). In
the case of NGC\,4088 and NGC\,6217 it was necessary to eliminate the
170$\mu$m measurements, which were clearly discrepant and probably had
calibration problems, and only retain longer wavelength measurements
(180$\mu$m and higher ISO measurements). Only for one of the galaxies
in the sample, NGC\,5860, a one temperature model was the best fit to
the infrared data (Calzetti et al. 2000).

The results of the fits to the far infrared data are presented in
Table~\ref{tabfir}. This table gives the infrared fluxes, integrated between
8$\mu$m and 1~mm, obtained from these fits, as well as the values extrapolated
based on IRAS measurements alone (F(IR) calculated using the expression from
Sanders \& Mirabel 1996). We also give the temperatures of the warm and cold
components, as well as their fractional contribution to the far infrared flux.

The total infrared emission derived from our fits has been compared to
the total infrared emission derived from the extrapolated IRAS
measurements.  In general we find that the discrepancy between the two
numbers is relatively small, of the order of 25\%, with a median ratio
F(IR)/F$_{IRAS}$(IR)=0.94.  Figure~2 shows that there is a clear
trend between F(IR)/F$_{IRAS}$(IR) to increase for large F100/F60
(ratio between the IRAS 100$\mu$m and 60$\mu$m fluxes). The relation
between the two quantities can be expressed by the linear fit

\begin{equation}
\log F(IR)/F_{IRAS}(IR) = -0.08 \pm 0.01 + (0.19 \pm 0.04) \log F100/F60
\end{equation}

Since F100/F60 is a temperature indicator, the above relation shows
that the IRAS extrapolation overpredicts the infrared emission for
warmer sources and underpredicts it for the cooler ones. This result
is in line with those found by Dale et al. (2001), which indicates that
the Sanders \& Mirabel (1996) relation can give results that are up
to $\sim$25\% deviant.  Dale et al. (2001) also point out that the ratio
F60/F100 produces the tightest correlation with other infrared
measurements, making it the ideal choice to parameterize the
correction for F$_{IR}$(IRAS). We use equation~(1) to correct the
total infrared fluxes for the 12 galaxies ($\sim$29\%~ of the sample)
for which IRAS but not ISO measurements are available.

\subsection{ULIGs, LBGs and SCUBA Sources}

In order to compare the local galaxies with their high redshift
counterparts we retrieved data from the literature for Arp\,220,
another 4 ULIGS, 13 LBGs, and 31 SCUBA sources (the latter in the
redshift range 2--3.5), for which both rest frame UV and rest frame
$\sim200\mu$m measurements were available.

For Arp\,220 and the 4 ULIGs (IC\,883, Mrk\,273, IRAS\,15250$+$3609 and
IRAS\,19254--7245) UV data was obtained from Goldader et al. (2002),
while the 205$\mu$m and FIR data came from ISO and IRAS observations
published by Klaas et al. (2001). We also obtained radio 6~cm data for
Arp\,220 (Becker, White \& Edwards 1991). In the case of the LBGs, their UV
emission is well known, from selection, but the majority
of these sources are not detected in submm observations. Here we use the
sample of LBGs from Chapman et al. (2000), which comprises 13 galaxies
with z$\sim$3, observed at 850$\mu$m with SCUBA. Only one of these sources
was detected (W-MMD\,11), while the remaining ones had only upper limits.
We assume that for these galaxies the upper limit of SFR(BOL$_{SB}$), the
star formation rate calculated using the bolometric luminosity, is equal
to the SFR(850$\mu$m) upper limit given by Chapman et al. (2000).
We also assume the 1$\sigma$ r.m.s. value to be the 200$\mu$m upper limit.
                                                                                
A situation opposite to the LBGs happens for SCUBA sources, which are
detected in the submm, but not always have optical counterparts or
spectroscopic redshifts. Using the compilations from Chapman et
al. (2003, 2005) we selected 31 SCUBA sources, from an initial sample
of 73, with spectroscopic redshifts z$>2$. These 31 galaxies do not
include sources classified as AGN (Aretxaga et al. 2005; Chapman et
al. 2005), although we include one source with a composite AGN/Starburst
characteristics.  The measurements obtained from these
papers are the broad band $\cal{R}_{AB}$ magnitudes, which were converted to
rest frame UV fluxes, radio 20~cm measurements, as well as bolometric
luminosities calculated fitting SEDs to the 850$\mu$m and radio fluxes
(Chapman et al. 2003, 2005).  These values are used to calculate
SFR(BOL$_{SB}$), and L$_{\lambda}$(205$\mu$m)/L$_{\lambda}$(UV) (we define 
L$_{\lambda}$(205$\mu$m) as the restframe luminosity measured with
SCUBA for uniformity with the other galaxy samples, although the restframe
wavelength is in the range 210-260$\mu$m).

\section{Calibrations of Star Formation Rates}

The UV, H$\alpha$ and IR star formation rates were calculated using the
calibrations presented by Kennicutt (1998):

\begin{eqnarray}
SFR(H \alpha) = 7.9 \times 10^{-42} L(H \alpha)&[erg~s^{-1}]&\\
SFR(UV) = 1.4 \times 10^{-28} L_{\nu}(1600)&[erg~s^{-1}~Hz^{-1}]&\\
SFR(IR) = 4.5 \times 10^{-44} L(IR)              &[erg~ s^{-1}]&
\end{eqnarray}

Where all the SFRs are in solar masses per year calculated for a Salpeter IMF
between 0.1 and 100 M$_{\odot}$.  The UV and H$\alpha$ luminosities
were corrected for foreground Galactic extinction, but not for internal
extinction. The SFR(IR) is a correct estimator of the star formation rate in
a galaxy only for large dust optical depths and young star dominated
spectral energy distributions (Kennicutt 1998).
In the case of the radio we followed the same approach used by
Condon \& Yin (1990) and Condon (1992) to derive a new calibration,
described below.

First, we assume that the non-thermal radio emission from the Milky Way is
entirely due to supernova remnants. This flux is L(408~MHz)$\sim6\times10^{21}$
W~Hz$^{-1}$ (Berkhuijsen 1984) with a spectral index $\nu^{-0.8}$. For the
supernova rate we use $\nu_{SN}\sim0.023$yr$^{-1}$ (Tammann 1982; Tammann,
Loeffler, \& Schroeder 1994). This rate corresponds only to core collapse
supernovae and was calculated assuming that the Milky Way is an Sc galaxy.
Notice that the percentage of core-collapse to Type Ia supernovae would
decrease by a factor of $\sim$10\% if the Milky Way was assumed to be an Sb
galaxy. These numbers also carry some uncertainties due to biases in the
detection of supernovae and the corrections that are applied in the
determination of supernova rates (Cappellaro et al. 1997; Cappellaro,
Evans, \& Turatto 1999; and references therein). Type Ia supernovae are
not taken into account in the SFR calculations because they are traditionally
related to the old stellar population. However, recent results suggest that
this may not be true (Mannucci et al. 2005; Scannapieco \& Bildsten 2005),
which may require a future revision of this calibration.

Following these assumptions, the non-thermal radio luminosity due to
supernovae can be calculated from
$L_{\nu}(N) \sim 1.3\times10^{23} \nu^{-0.8} \nu_{SN}$, in units of W~Hz$^{-1}$,
the frequency ($\nu$) is in units of GHz and the supernova rate ($\nu_{SN}$)
in units of yr$^{-1}$. The
conversion between this value and the star formation rate is done in the same
way Kennicutt (1998) derived the IR calibration. Using a Starburst 99
(Leitherer et al. 1999) model for a galaxy with continuous star formation
rate of 1~M$_{\odot}$~yr$^{-1}$, with a Salpeter IMF, solar metallicity,
lower and upper mass cutoff 1 and 100 M$_{\odot}$, respectively, we find that
the average supernova rate reaches an asymptotic value of
$\nu_{SN}\sim0.0168$yr$^{-1}$ around 100~Myr. Converting the star formation
rate to a lower mass cutoff of 0.1~M$_{\odot}$ and using the above relation
between non-thermal radio emission and the supernova rate, we get:

\begin{equation}
L_{\nu}(N) \sim 8.55 \times 10^{20} \nu^{-0.8} SFR
\end{equation}

\noindent
where SFR is the rate of stars in the mass range 0.1-100~M$_{\odot}$ being
formed, in units of M$_{\odot}$~yr$^{-1}$, $\nu$ is the frequency of the radio
observations, in GHz, and L$_{\nu}(N)$ is the non-thermal radio luminosity at this
frequency, in units of W~Hz$^{-1}$.

The contribution from thermal emission to the radio continuum is taken from
Condon (1992), where we used the new H$\alpha$ to SFR conversion by Kennicutt
(1998), thus extending the lower mass cutoff of the stellar IMF to
0.1~M$_{\odot}$.

\begin{equation}
L_{\nu}(T)\sim 1.6\times 10^{20} \nu^{-0.1} SFR
\end{equation}

The total SFR at radio wavelengths is therefore calculated by simply
summing up the appropriate thermal and non-thermal contributions at a
given frequency

\begin{equation}
SFR(RADIO) = L_{\nu} \times 10^{-20}/(8.55 \nu^{-0.8} + 1.6 \nu^{-0.1}) [W Hz^{-1}]
\end{equation}

\noindent
which at 1.4~GHz corresponds to SFR$ = 1.24\times10^{-21}$ L$_{\nu}$(1.4 GHz).
Notice that, although the radio emission has the
advantage of being independent of reddening, it is not as direct a
tracer of star formation as UV and H$\alpha$, given the assumptions
included in the derivation of the non--thermal portion.

Finally, we also define a new multiwavelength SFR indicator, by
combining the emission from the UV, B and IR bands, which we call
SFR(BOL$_{SB}$). This quantity is calculated by integrating the SED between
the UV and B part of the spectrum, and adding this quantity to L$_{IR}$.
The specifier BOL$_{SB}$ refers to the fact that we are essentially computing
the bolometric luminosity of the starburst, as these blue bands are
dominated by the emission from the young massive stars of the
starburst itself (the B band data of the galaxies in the sample is
listed in Paper~I). This is particularly important for a sample that spans
a wide range of luminosities, as the less luminous objects tend to be less
dust opaque and thus to have a higher fraction of the stellar light coming out
directly in the UV and B. The corresponding SFRs are calculated using the
IR calibration from Kennicutt (1998). This is an extension of the
SFR(UV$+$IR) introduced by Wang \& Heckman (1996) and Heckman et
al. (1998) for normal star--forming and starburst galaxies,
respectively.  BOL$_{SB}$ is a more accurate estimator of SFR than
L$_{IR}$, since in our galaxies a non--negligible fraction of the
stellar light emerges directly in the UV and B bands, unabsorbed by
dust; hence L$_{IR}$ alone is not a good approximation to the bolometric
luminosity, especially for the less bright and less dusty galaxies.
Incidentally, although SFR(BOL$_{SB}$) also includes the
B band flux, it generally gives values that are on average very
similar to SFR(UV$+$IR), with a spread smaller than 10\% around the median.

\section{Comparison of Star Formation Rates}
                                                                                
\subsection{Matched Apertures}
 
Given the small aperture with which the UV data were obtained, we compare
their SFRs only with H$\alpha$ and radio (8.4~GHz) ones measured inside a
matching aperture. In Figure~3 we show the ratios SFR(H$\alpha$)/SFR(UV)
(top panel) and SFR(UV)/SFR(8.4~GHz) (bottom left panel) as a function of
SFR(8.4~GHz). In both cases the SFR derived from UV is lower than the one
derived from the other indicator, clearly showing the effects of extinction.
Of particular interest is the large scatter in the SFR(UV)/SFR(8.4~GHz)
plot, which clearly shows that dust extinction has a strong effect on
measurements of SFR at short wavelengths. A similar result is seen in
the SFR(H$\alpha$)/SFR(8.4~GHz) plot (bottom left panel). This highlights the
importance of extinction corrections, since in some cases the uncorrected UV
and H$\alpha$ measurements can underpredict the SFR by a factor as high as 100.
Nevertheless, the SFR(H$\alpha$)/SFR(UV) ratio remains strongly correlated,
suggesting that both wavebands probe comparably low extinction regions.
We stress the importance of subtracting the [NII] contribution from the flux
in H$\alpha$ images, as [NII] contamination becomes increasingly important
in high metallicity systems (Storchi-Bergmann, Calzetti, \& Kinney 1994),
where the [NII] emission can contribute to as much as half of the total flux.
 
\subsection{Integrated Apertures}
 
Emission integrated over the entire body of the galaxies is available
at H$\alpha$, IR, and radio wavelengths. Here we compare global SFRs
among the different wavebands (Figure 4). We generally find a
good agreement between the different indicators, with a few deviations.
In the case of H$\alpha$, the SFR usually is underestimated because of
dust extinction. We find that the amount of extinction increases for
high luminosity sources. This result represents an independent confirmation
of those from Wang \& Heckman (1996), Heckman et al. (1998),
Sullivan et al. (2001), Martin et al. (2005) who find a tred for more
opaque galaxies to have higher star formation rates (see Section~5 for further
discussion on this subject). Figure~4 gives further support to this
interpretation, where we can see that using uncorrected H$\alpha$ fluxes
can underestimate the star formation rate by a factor of 10 or larger
in the most luminous sources. This result agrees with those from Cram et
al. (1998), Sullivan et al. (2001) and Afonso et al. (2003).
 
Another noticeable deviation is found in the comparison of SFRs in different
radio wavelengths, where we can see that the 8.4~GHz generally gives lower
values than 1.4~GHz. We believe that this is due to limitations of our
observations, since the 8.4~GHz images had a beamsize of $\sim$3\arcsec\ and
maximum field of view of 180\arcsec, thus missing some of the faint
diffuse 8.4~GHz emission in the more extended sources. This was not an issue
for the 1.4~GHz measurements.
                                                                                
\subsection{New Radio SFR calibration}
                 
One of the most important results obtained from the comparison of
the integrated star formation indicators is shows in the top panels of
Figure~5. This Figure shows the discrepancy between SFR(BOL$_{SB}$) and
SFR(Radio), both 4.89~GHz and 1.4~GHz, in the sense that the values
determined from radio measurements tends to be higher than the ones obtained
from the bolometric luminosity by a factor of $\sim$2. This difference is
similar to the one found by Condon, Cotton \& Broderick (2002), and
Bell (2003). 

This disagreement could be the result of an overestimation of the radio
relation, or to an underestimation of the BOL$_{SB}$ (infrared) relation.
Taking into account the fact that the infrared SFR relation is a more direct
calibration than the radio one, and that it has been tested against
calibrations at other wavebands (e.g. Kewley et al. 2004, Rosa-Gonz\'alez
et al. 2002), we attribute the observed disagreement to the radio calibration.
Notice however, that an opposite result was
presented by Cappellaro et al. (1999), who found that the Far-Infrared
luminosity does not correlate with supernova rates in nearby galaxies.
Although this result seems to indicate that infrared is not a universal
measurement of SFR, their measurements were biased towards normal low
luminosity quiescent galaxies, and cannot be considered representative of
the sources studied in this paper.
Furthermore, since we used BOL$_{SB}$ instead of L(IR), we removed most of
systematic effects in the infrared calibration. 
                       
Here we propose a correction to the radio SFR relation, based on the
comparison between the 1.4~GHz and BOL$_{SB}$ results. We find that the
median ratio SFR(BOL$_{SB}$)/SFR(1.4~GHz) is 0.48, which gives us the
corrected SFR(1.4~GHz) relation,
 
\begin{equation}
SFR = 6.2\times10^{-22} L_{\nu}(1.4 GHz)    [W Hz^{-1}]
\end{equation}
                                                                                
\noindent
A comparison between this new relation and the one from Yun, Reddy \& Condon
(2001), who derived SFR(1.4~GHz) based on the comparison between the
integrated radio and infrared luminosity functions, shows a good agreement.
We also find a good agreement when comparing our result with the one from
Bell (2003). However, given the luminosity range of our galaxies we are
not able to see the deviation from linearity seen by Bell (2003) in the low
luminosity range. Using the SFR(IR) calibration from
Kennicutt (1998) and the radio-FIR correlation given by Yun et al. (2001)
we get an SFR(1.4~GHz) relation in agreement with equation~8 within an
uncertainty of 20\%. The uncertainty is mostly driven by the assumption
used to convert L(40-120$\mu$m) to L(IR). A variation of about 20\%
is typical of what is found in galaxies (Dale et al. 2001).

Next we determine a new relation for the non-thermal part of the radio
SFR calibration (Equation~5). Since the non-thermal part of the
relation is based on indirect assumptions, which depend mainly on the
supernova rate of the Galaxy and their average non-thermal flux, it is
subject to large uncertainties.  For instance, Condon (1992) pointed
out the discrepancy between the predicted and observed supernova rate
of the Galaxy. Another potential problem is that the measured 408~MHz
flux of the Milky Way carries large uncertainties. Conversely, we
assume that the uncertainties in the thermal part of the calibration
(Equation~6) are insignificant when compared to the non-thermal ones,
since that relation is derived directly from SFR(H$\alpha$)
(Kennicutt 1998), which has been tested against other
calibrations. Using Equations~6 and 8 we find the following
non-thermal radio SFR relation:
 
\begin{equation}
L_{\nu}(N) \sim 1.92 \times 10^{21} \nu^{-0.8} SFR
\end{equation}
 
Using Equations 6 and 9 we can calculate the radio SFR relations at 4.89~GHz
and 8.46~GHz, given below:
 
\begin{eqnarray}
SFR = 1.4\times10^{-21} L_{\nu}(4.89 GHz)& [W Hz^{-1}]&\\
SFR = 2.0\times10^{-21} L_{\nu}(8.46 GHz)& [W Hz^{-1}]&
\end{eqnarray}
 
\noindent
A comparison between the new SFR(4.89~GHz) and SFR(1.4~GHz) with values
obtained from BOL$_{SB}$ (Figure~5 bottom panels) shows a very good agreement,
indicating that the new radio
relations are indeed more appropriate for the calculation of SFRs. We do not
find as good an agreement for 8.46~GHz, but this is related to observational
issues described above. It should also be noticed that Equation~10 has
a direct application to high-redshift
galaxies, where 20~cm observations correspond to restframe 6-7~cm for a
z$\sim$3 object.

\subsection{Extinction and other Effects Influencing SFRs}

The effects of extinction suggested by Figure~3 are quantified in
Figure~6.  This figure shows the color excess E(B-V) derived from each
pair of wavebands in Figure~3, requiring that any deviation of the SFR
ratios from unity is due to dust extinction. In the case of SFR(8.4~GHz)
we use the new calibration described above. For the UV, the color excess
is derived using the reddening curve of Calzetti et al. (2000). We find
that the different E(B-V) estimates have similar medians, albeit with a
wide range of values, implying that dust geometry has a secondary impact
(beyond that described by Calzetti 2001) on the SFR measurements at
different wavelengths for our galaxy sample. A comparison of E(B-V) values
calculated using UV/8.4~GHz and H$\alpha$/8.4~GHz, for galaxies with
measurements available in the 3 wavebands, show that they
are correlated. This indicates that the SFR differences relative to
SFR(8.4~GHz) are due to dust extiction. It is also worth recalling here
that although E(B-V)$_{UV}\sim$E(B-V)$_{H\alpha}$, the resulting
A$_{UV}\sim2.3$~mag$>$A$_{H\alpha}\sim1.5$~mag, thus explaining the
trend in Figure~3.

Another important result of Figure~6 is the large spread of the
extinction corrections among the galaxies in our sample, which have an
interquartile range of the order of 0.5~mag. This indicates that
assuming a single extinction value may work only on a statistical
sense, for large samples of galaxies with similar
characteristics. More detailed studies of this issue were performed by
Bell \& Kennicutt (2001), Rosa-Gonzalez, Terlevich, \& Terlevich
(2002), and Afonso et al. (2003).

Our SFR estimates are based on the assumption that we can use one
conversion from flux to SFR for all galaxies. Clearly, this will be
valid only in first order, as galaxy--to--galaxy variations of the
stellar Initial Mass Function (IMF), metallicity (Leitherer et
al. 1999), and star formation history (Sullivan, et al. 2004), in addition to
dust absorption of Lyman continuum photons before reprocessing into
H$\alpha$ light (Inoue 2001), and simplistic assumptions about the
conditions of nebular recombination (e.g. Charlot \& Longhetti 2001),
will all affect SFR measurements. Indeed, we believe these potential
galaxy--to--galaxy variations to contribute, together with dust
geometry variations, to the scatter in the data observed in
Figures~3 to 6.

As an example, we will discuss the impact of stellar IMF
variations. The part of the IMF of concern here is the high end, as
massive stars are those contributing to the UV emission and to the gas
ionization. From Leitherer et al. (1999), a stellar population with a
M$_{up}$=30~M$_{\odot}$ Salpeter IMF produces about 5 times less
ionizing photons and about twice less UV continuum flux than a stellar
population with a M$_{up}$=100~M$_{\odot}$ Salpeter IMF. Thus,
SFR(UV)/SFR(H$\alpha$) calculated for the two populations will differ
by about a factor of 2.5, without having to invoke any other effect
(e.g., dust geometry variations). SFR(radio) closely follows the
variation of SFR(H$\alpha$), being due to similar stars. If
interpreted as a difference due to dust, it would correspond to
E(B$-$V)$\sim$0.5~mag, roughly the range of our scatter (Figure~6, top
panel).  We note that this should be considered an extreme scenario,
since Elmegreen (2005, Cambridge Prooceedings) has collected evidence
that IMF variations are likely to be relatively small from galaxy to
galaxy, and from environment to environment. Metallicity variations
introduce even smaller effects than IMF variations on the
multiwavelength SFR determinations (Leitherer et al. 1999). 

\section{Comparison of the UV/IR Properties of Nearby and High-Redshift Galaxies}

As a first test, to see if there is any systematic difference between
our galaxies and the general population of nearby sources, we show
in the left panel of Figure~7 the plot of radio 1.4~GHz $\times$ FIR
(FIR is calculated as described in Sanders \& Mirabel 1996, using
only the 60 and 100$\mu$m fluxes). A comparison between the relation
obtained by Condon, Anderson \& Helou (1991), solid line, and the values
measured for our galaxies show a very good agreement, indicating that our
sample is representative of the population of nearby star--forming galaxies.

Another test performed with the data was to check if there is any difference
between the nearby galaxies and the SCUBA sources (Figure~7 right panel).
This was done by comparing the ratio of monochromatic IR to radio (6~cm)
emission of our galaxies, Arp\,220 and the SCUBA sources from Chapman et al.
(2003, 2005). We chose this radio frequency because it corresponds to
observed $\sim$20~cm at z$\sim$2.2, the median redshift of the SCUBA galaxies,
and is the frequency at which most of the deep radio surveys are
done. Given the uncertainty in the slopes of the radio spectrum of
these sources we do not try to apply any correction to the data, to
put all the measurements in the same rest frame frequency. This
figure shows that there is no strong correlation between
L$_{\lambda}$(205$\mu$m)/L$_{\nu}$(6~cm) and SFR(BOL$_{SB}$), apart from a
small decrease in the ratio for higher SFRs (higher IR luminosity). This is an
expected result, in line with those from Condon et al. (1991),
thus confirming that there are no problems with the sample. According to
Yun, Reddy \& Condon (2001), this slope as a function of luminosity is
due to the contribution from the general field populations to the
far-infrared emission in quiescently star--forming galaxies.

The star--forming galaxies in our sample show the same range in
IR$+$UV luminosity properties as those shown by the starburst galaxies
analyzed by Heckman et al. (1998). In particular, there is a trend for
more luminous galaxies to have a higher IR/UV luminosity ratio
(Figure~\ref{fig8}), indicating that more actively star--forming galaxies
also tend to be more dust opaque, a trend already noted in other samples
(Wang \& Heckman 1996, Heckman et al. 1998, Sullivan et al. 2001,
Hopkins et al. 2001; Martin et al. 2005). We compare in Figure~8
the best fitting relation obtained from our data (dotted line) with the
one obtained when fitting the starbursts data from Heckman et al. (1998),
using  L(IR)$+\lambda$L$_{\lambda}$(UV) as independent variable (solid line).
We can see that both samples produce very similar fits, which also is
consistent with the trend seen in the UV selected star forming galaxies
from Martin et al. (2005), given the large uncertainties in our sample
and the different choice of IR luminosities in the Martin et al. (2005).
The only difference is that our galaxies at the bright end tend to have on
average slightly lower, by a factor of a few,
L(IR)/L$_{\lambda}$(UV) ratio at the same total IR$+$UV luminosity.
This difference can be explained if not all the observed UV emission is
associated with current star formation, as observed in NGC5194
(Calzetti et al. 2005), and/or if the starburst dust geometry is not
readily applicable to star--forming galaxies; both effects can boost
the UV emission relative to the infrared (Calzetti 2001, Buat et
al. 2002). As far as SFR(BOL$_{SB}$) is concerned, the impact on this value
of the UV emission unrelated to current star formation will be minimal
as L$_{IR}$ provides most of the bolometric light for our brightest normal 
star forming galaxies.
   
Conversely, our sample galaxies do not show any clear trend for the ratio
L$_{\lambda}$(205~$\mu$ m)/L$_{\lambda}$(UV) as a function of SFR(BOL$_{SB}$)
(equivalent to the IR$+$UV luminosity, Figure~9 left; Meurer et
al. 1999). The trend is absent even if galaxies with more than 50\% of
the star formation {\it outside} the STIS aperture are excluded from
the plot, to mitigate the aperture mismatch between the UV and IR data
(Figure~9, right); the selection is performed by excluding galaxies with
more than 50\% of their radio or H$\alpha$ emission outside the area covered
by the STIS aperture, under the assumption that the 3.6~cm emission is
a good tracer of unobscured star formation. The range of
L$_{\lambda}$(205~$\mu$m)/L$_{\lambda}$(UV) ratios covered
by the local star--forming sample
is typically between 0.6 and 15, with a couple of outliers (NGC\,3079
at the top and TOL\,1924$-$416 at the bottom).
   
The absence of a trend for the L$_{\lambda}$(205~$\mu$m)/L$_{\lambda}$(UV)
ratio of our star--forming galaxies can be understood by recalling that the
$\sim200~\mu$m emission is located in the Rayleigh--Jeans tail of the
FIR emission, and is most sensitive to the emission from dust at
temperatures $<$20~K (cirrus). This dust can be heated by the non-ionizing
(non--star--forming) stellar populations in the host galaxies, and
does not necessarily correlates with the star formation.
   
Interestingly, the scatter plot of Figure~9, right panel, changes to a mild
trend of higher L$_{\lambda}$(205~$\mu$ m)/L$_{\lambda}$(UV) ratios for higher
SFRs when data on the ULIGs are added to the Figure. This likely reflects the
extreme nature of the ULIGs, with large bolometric FIR fluxes and very faint
UV emission. Thus, even when observing the monochromatic FIR emission
in the Rayleigh--Jean tail some trend of higher opacities for larger
SFRs is preserved, albeit with a much larger scatter than when using
the bolometric FIR emission. Again, this mild trend is highly sensitive
to contribution to the L$_{\lambda}$(205~$\mu$m) by cirrus emission, and
can only be observed when the complete range of SFRs in the local
Universe, from mild star--forming galaxies to ULIGs, is
included. Including the 31 SCUBA sources with available restframe UV
data amplifies the trend towards larger, by 2--3 order of magnitude,
L$_{\lambda}$(205~$\mu$m)/L$_{\lambda}$(UV) values as the SFR increased by
roughly 2 orders of magnitude (Figure~9). In contrast, the lonely
SCUBA--detected LBGs, and the 12 upper limits are not incompatible with the
L$_{\lambda}$(205~$\mu$m)/L$_{\lambda}$(UV) range of local galaxies and,
if anything, they may
have lower ratios, much more similar to the dust--poor TOL\,1924--416
(Figure~9).

For the 12 LBGs undetected by SCUBA (Chapman et al. 2000), we now
attempt to use the local star--forming galaxies
L$_{\lambda}$(205~$\mu$ m)/L$_{\lambda}$(UV) ratio range to predict what their
SCUBA fluxes could be. The undetected LBGs cluster around observer--frame
magnitudes $\cal{R}_{AB}\sim$24 (Chapman et al. 2000). For our observed
range of L$_{\lambda}$(205~$\mu$m)/L$_{\lambda}$(UV) ratios (Figure~9, right),
the LBGs should then have monochromatic far infrared fluxes in the range
0.7--17~mJy in the SCUBA band. Thus, at least some of those 12 LBGs
should have been detected, while none was at the r.m.s. sensitivity
level of $\sim$1~mJy.
   
This discrepancy between expectations and reality calls into question
the applicability of the local flux ratios to the high--redshift case,
since the Rayleigh--Jeans tail of the FIR emission receives a
potentially large contribution from the dust heating by non-ionizing
stellar populations in local galaxies. The case of TOL\,1924$-$416 is
illuminating in this respect. This Blue Compact Galaxy shows a UV and
optical spectrum typical of a young--starburst--dominated galaxy, with
a CIV(1550~\AA) P-Cygni profile, a large H$\alpha$ line emission
equivalent width ($\sim$180~\AA), and a very blue UV--optical spectral
energy distribution (Kinney et al. 1993, Storchi-Bergmann, Kinney, \&
Challis 1995); it shows a large
gas--to--dust ratio, that has been suggested due to the absence of a
non-ionizing stellar population (Gondhalekar et al. 1986). The latter is in
line with its very hot FIR SED, with L(IR)/L$_{\lambda}$(UV)=1.3, but
L$_{\lambda}$(205~$\mu$ m)/L$_{\lambda}$(UV)=0.025 (Calzetti et al. 2000,
see also Figure~9). If the L$_{\lambda}$(205~$\mu$m)/L$_{\lambda}$(UV) ratio
observed in TOL\,1924$-$416 is more typical of LBGs, the expected SCUBA fluxes
for these galaxies would be around 0.03~mJy, thus explaining why they are
undetected.
   
It is worth remarking that the absence of a non-ionizing population
contributing to the heating of the dust does not preclude the LBGs
from following the general correlation between L(IR)/L$_{\lambda}$(UV)
versus UV colors (Meurer et al. 1999, Calzetti 2001) or versus total
IR$+$UV luminosity (Heckman et al. 1998) observed in local starburst
galaxies (Adelberger \& Steidel 2000). In fact, TOL\,1924$-$416 does follow
those correlations (Calzetti 2001), as they involve the bolometric FIR
emission, rather than a monochromatic one. The dust emission at $\sim$200~$\mu$m
usually represents a small fraction, 5\% or less, of the total
FIR emission from UV--selected starburst galaxies (Calzetti et
al. 2000), thus explaining its small impact on correlations involving
the bolometric FIR emission.

\section{Summary}

We presented a comparison between a set of 4 star formation
indicators, spanning the wavelength range from UV to radio.  We
discuss the effect of dust extinction and calibration in their
estimates. We find that, as previously pointed out, dust extinction
has a strong impact at lower wavelengths, where it can severely
underestimate the SFR by up to two orders of magnitude in individual
objects. However, we still find that UV and H$\alpha$ are well
correlated, indicating that these two measurements come from similar
region, although with different optical depths. The amount of extinction
varies significantly from galaxy to galaxy in our sample, with a
spread larger than 0.5 magnitudes in color excess E(B--V). Other factor that
can affect the determination of SFRs are also discussed. A
comparison between the bolometric and radio SFRs shows that the
latter calibration overestimates the star formation rates. Based on
the assumption that this discrepancy is due to uncertainties in the
non-thermal part of the radio calibration, we provide a new
calibration for SFR(radio). In particular our new SFR(6~cm) calibration
has direct application for high-redshift objects, where the observed 20~cm
fluxes correspond to restframe 6-7~cm for a z$\sim$3 galaxy.

We also compared the ratio L$_{\lambda}$(205$\mu$m)/L$_{\lambda}$(UV)
between our normal galaxy sample and higher redshift galaxies. We find a trend
for higher ratios as the star formation increases, thus suggesting that a
fraction of the L$_{\lambda}$(205$\mu$m) emission is heated by the star
forming population and the L$_{\lambda}$(205$\mu$m)/L$_{\lambda}$(UV) ratio
is still measuring dust opacity. The SCUBA sources occupy a locus in
the L$_{\lambda}$(205$\mu$m)/L$_{\lambda}$(UV)--versus--SFR plane that is
at the high end of that occupied by the ULIGs, suggesting more extreme
star formation conditions that in local ULIGs. LBGs may instead
resemble the local star--forming galaxies, and could even be located at
lower L$_{\lambda}$(205$\mu$m)/L$_{\lambda}$(UV) ratios, although any accurate
comparison is prevented by the lack of detections in the sub-mm for LBGs.
Consistency checks indicate that LBGs may indeed resemble the local metal-poor
and dust-poor starburst galaxies, rather than the average local population.

\acknowledgements
 
This work was partially supported by the NASA grants HST-GO-8721, and NAG5-8426.
The National Radio Astronomy Observatory is a facility of the National
Science Foundation, operated under cooperative agreement by Associated
Universities, Inc. This research made use of the NASA/IPAC Extragalactic
Database (NED), which is operated by the Jet Propulsion Laboratory, Caltech,
under contract with NASA. HRS would like to acknowledge the NRAO Jansky
Fellowship program for support during most of the stages of this project.
HRS would also like to thank the Spitzer Science Center,
and the Space Telescope Science Institute visitor programs for their support.
The UV observations were obtained with the NASA/ESA Hubble Space Telescope at
the Space Telescope Science Institute, which is operated by the Association of
Universities for Research in Astronomy, Inc., under NASA contract NAS5-26555.
Basic research at the US Naval Research Laboratory is supported by the Office
of Naval Research. We would like to thank the referee for comments that helped
us improve this paper.

\clearpage
\begin{deluxetable}{lrrrrrrrrrr}
\tabletypesize{\scriptsize}
\tablewidth{0pc}
\tablecaption{Far Infrared Fluxes and Fit Results}
\tablehead{
\colhead{Name}&
\colhead{F$_{IR}$}&
\colhead{F$_{IR}$(IRAS)}&&
\colhead{F(60$\mu$m)}&
\colhead{F(100$\mu$m)}&&
\colhead{T$_w$}&
\colhead{T$_c$}&
\colhead{f$_w$}&
\colhead{f$_c$}\\
\cline{2-3} \cline {5-6} \cline{8-9}\\
\colhead{}&
\multicolumn{2}{c}{(10$^{-11}$~erg~cm$^{-2}$~s$^{-1}$)}&&
\multicolumn{2}{c}{(Jy)}&&
\multicolumn{2}{c}{(K)}&
\colhead{}&
\colhead{}\\
\colhead{(1)}&
\colhead{(2)}&
\colhead{(3)}&&
\colhead{(4)}&
\colhead{(5)}&&
\colhead{(6)}&
\colhead{(7)}&
\colhead{(8)}&
\colhead{(9)}}
\startdata
ESO\,350-G\,38 &\nodata& 72.4 &&    6.48 &   5.01&&\nodata&\nodata&\nodata&\nodata \\
NGC\,232       &89.1   & 97.6 &&   10.04 &  18.34&&46     &23     &0.45   &   0.55 \\
MRK\,555       &43.4   & 47.2 &&    4.22 &   8.68&&33     &18     &0.71   &   0.29 \\
IC\,1586       &9.7    & 12.4 &&    0.96 &   1.69&&54     &23     &0.40   &   0.60 \\
NGC\,337       &\nodata& 81.0 &&    8.35 &  17.11&&\nodata&\nodata&\nodata&\nodata \\
IC\,1623       &183.0  & 209.2&&   22.58 &  30.37&&49     &24     &0.57   &   0.43 \\
NGC\,1155      &\nodata& 27.1 &&    2.45 &   4.60&&\nodata&\nodata&\nodata&\nodata \\
UGC\,2982      &80.8   & 89.8 &&    8.35 &  16.89&&41     &26     &0.31   &   0.69 \\
NGC\,1569      &\nodata& 381.1&&   45.41 &  47.29&&\nodata&\nodata&\nodata&\nodata \\
NGC\,1614      &260.6  & 311.5&&   32.31 &  32.69&&49     &26     &0.68   &   0.32 \\
NGC\,1667      &69.3   & 70.9 &&    5.95 &  14.73&&49     &23     &0.29   &   0.71 \\
NGC\,1672      &\nodata& 357.0&&   32.96 &  69.89&&\nodata&\nodata&\nodata&\nodata \\
NGC\,1741      &\nodata& 36.7 &&    3.92 &   5.84&&\nodata&\nodata&\nodata&\nodata \\
NGC\,3079      &496.1  & 425.3&&   44.50 &  89.22&&38     &16     &0.53   &   0.47 \\
NGC\,3690      &\nodata& 982.6&&  103.70 & 107.40&&\nodata&\nodata&\nodata&\nodata \\
NGC\,4088      &227.6  & 226.2&&   19.88 &  54.47&&37     &22     &0.35   &   0.65 \\
NGC\,4100      &96.7   & 96.4 &&    8.10 &  21.72&&48     &23     &0.23   &   0.77 \\
NGC\,4214      &\nodata& 172.0&&   17.87 &  29.04&&\nodata&\nodata&\nodata&\nodata \\
NGC\,4861      &\nodata& 20.4 &&    1.97 &   2.46&&\nodata&\nodata&\nodata&\nodata \\
NGC\,5054      &132.3  & 130.0&&   11.60 &  26.21&&35     &18     &0.56   &   0.44 \\
NGC\,5161      &\nodata& 29.9 &&    2.18 &   7.24&&\nodata&\nodata&\nodata&\nodata \\
NGC\,5383      &\nodata& 62.3 &&    4.89 &  13.70&&\nodata&\nodata&\nodata&\nodata \\
MRK\,799       &95.8   & 112.1&&   10.41 &  19.47&&33     &18     &0.83   &   0.17 \\
NGC\,5669      &23.9   & 20.4 &&    1.66 &   5.19&&39     &19     &0.31   &   0.69 \\
NGC\,5676      &128.2  & 126.8&&    9.64 &  30.66&&39     &23     &0.19   &   0.81 \\
NGC\,5713      &184.4  & 207.8&&   19.82 &  36.20&&37     &22     &0.60   &   0.40 \\
NGC\,5860      &16.3   & 18.0 &&    1.64 &   3.02&&32     &\nodata&1.00   &   0.00 \\
NGC\,6090      &58.8   & 63.7 &&    6.66 &   8.94&&49     &23     &0.59   &   0.41 \\
NGC\,6217      &105.6  & 112.4&&   10.83 &  19.33&&43     &21     &0.51   &   0.49 \\
NGC\,6643      &127.9  & 128.0&&    9.38 &  30.69&&47     &24     &0.11   &   0.89 \\
UGC\,11284     &88.0   & 83.8 &&    8.25 &  15.18&&46     &18     &0.50   &   0.50 \\
NGC\,6753      &109.4  & 114.4&&    9.43 &  27.36&&30     &23     &0.52   &   0.48 \\
TOL\,1924-416  &11.1   & 15.3 &&    1.69 &   1.01&&50     &31     &0.99   &   0.01 \\
NGC\,6810      &185.9  & 203.9&&   17.79 &  34.50&&57     &26     &0.27   &   0.73 \\
ESO\,400-G\,43 &\nodata& 14.7 &&    1.59 &   1.58&&\nodata&\nodata&\nodata&\nodata \\
NGC\,7496      &84.3   & 90.6 &&    8.46 &  15.55&&49     &23     &0.42   &   0.58 \\
NGC\,7552      &664.6  & 701.6&&   72.03 & 101.50&&45     &19     &0.62   &   0.38 \\
MRK\,323       &37.1   & 38.8 &&    3.16 &   7.91&&32     &19     &0.63   &   0.37 \\
NGC\,7673      &42.0   & 43.3 &&    4.91 &   6.89&&43     &20     &0.67   &   0.33 \\
NGC\,7714      &92.6   & 106.6&&   10.36 &  11.51&&55     &25     &0.62   &   0.38 \\
MRK\,332       &47.0   & 54.2 &&    4.87 &   9.49&&34     &21     &0.74   &   0.26 \\
\enddata
\tablenotetext{.}{Column 1: galaxy name; columns 2 and 3: the far infrared
fluxes (8$\mu$m to 1mm), calculated using all infrared measurements available,
or only IRAS measurements, respectively; columns 4 and 5: 60$\mu$m and
100$\mu$m fluxes; columns 6 and 7: the fitted warm and cold temperatures;
columns 8 and 9: the fraction of the FIR flux due to the warm and cold
components.}
\label{tabfir}
\end{deluxetable}

\clearpage
\begin{deluxetable}{lrrrrrrrr}
\tabletypesize{\scriptsize}
\tablewidth{0pc}
\tablecaption{Integrated Star Formation Rates}
\tablehead{
\colhead{Name}&
\colhead{UV}&
\colhead{H$\alpha$}&
\colhead{H$\alpha^{cor}$}&
\colhead{8.46~GHz}&
\colhead{4.89~GHz}&
\colhead{1.4~GHz}&
\colhead{F$_{IR}$}&
\colhead{BOL$_{SB}$}\\
\cline{2-9}\\
\colhead{}&
\multicolumn{8}{c}{(M$_{\sun}$ yr$^{-1}$)}\\
\colhead{(1)}&
\colhead{(2)}&
\colhead{(3)}&
\colhead{(4)}&
\colhead{(5)}&
\colhead{(6)}&
\colhead{(7)}&
\colhead{(8)}&
\colhead{(9)}}
\startdata
ESO\,350-G\,38&\nodata &  21.57 &  17.59 &  28.63 &  32.20 &  28.24 &   24.62 &  25.85 \\
NGC\,232      &\nodata &   3.16 &   1.61 &  50.14 & 141.80 &  74.70 &   35.98 &  37.30 \\
MRK\,555      &   0.59 &   2.72 &   2.00 &   6.77 &   9.79 &  18.88 &    6.78 &   8.09 \\
IC\,1586      &\nodata &   1.37 &   1.37 &   4.56 &\nodata &   8.15 &    3.11 &   3.66 \\
NGC\,337      &\nodata &   1.56 &   1.18 &   1.65 &   6.01 &   7.29 &    1.77 &   2.11 \\
IC\,1623      &   5.37 &\nodata &\nodata & 135.30 & 197.00 & 249.00 &   60.06 &  63.66 \\
NGC\,1155     &\nodata &\nodata &\nodata &   3.28 &\nodata &   5.20 &    5.07 &   5.28 \\
UGC\,2982     &\nodata &\nodata &\nodata &  38.15 &  44.16 &  70.94 &   20.32 &  20.32 \\
NGC\,1569     &   0.03 &   0.12 &   0.12 &   0.03 &   0.17 &   0.13 &    0.05 &   0.05 \\
NGC\,1614     &\nodata &   3.94 &   3.94 &  67.88 &  78.28 &  83.60 &   51.67 &  52.31 \\
NGC\,1667     &   0.69 &   3.48 &   1.88 &  26.52 &  50.73 &  42.43 &   12.46 &  13.38 \\
NGC\,1672     &   0.10 &   0.35 &   0.35 &\nodata &   7.65 &  14.69 &    3.82 &   4.37 \\
NGC\,1741     &   2.08 &   3.68 &   3.24 &   6.84 &   6.13 &  13.70 &    5.19 &   6.51 \\
NGC\,3079     &   0.02 &\nodata &\nodata &  21.21 &  42.61 &  49.81 &   10.51 &  10.62 \\
NGC\,3690     &\nodata &  14.60 &  14.60 & 165.90 & 165.70 & 182.30 &   86.59 &  86.91 \\
NGC\,4088     &   0.02 &   1.71 &   1.32 &   2.72 &   6.18 &   7.81 &    3.35 &   3.47 \\
NGC\,4100     &\nodata &\nodata &\nodata &   3.21 &\nodata &   2.26 &    1.42 &   1.65 \\
NGC\,4214     &   0.03 &   0.10 &   0.09 &   0.11 &   0.12 &   0.07 &    0.11 &   0.15 \\
NGC\,4861     &   0.53 &   0.61 &   0.61 &   0.78 &   0.84 &   0.71 &    0.41 &   0.65 \\
NGC\,5054     &\nodata &\nodata &\nodata &   5.47 &  11.65 &  10.38 &    5.02 &   5.72 \\
NGC\,5161     &\nodata &   1.08 &   0.83 &   0.52 &\nodata &   1.97 &    1.71 &   2.55 \\
NGC\,5383     &   0.19 &   5.91 &   5.91 &   3.57 &   5.01 &   6.81 &    4.53 &   5.50 \\
MRK\,799      &   0.12 &   2.16 &   1.43 &  10.74 &  16.39 &  18.43 &    8.64 &   8.89 \\
NGC\,5669     &   0.08 &\nodata &\nodata &\nodata &\nodata &   1.83 &    0.76 &   0.98 \\
NGC\,5676     &   0.06 &   2.70 &   2.70 &   7.77 &  14.43 &  21.96 &    7.77 &   7.90 \\
NGC\,5713     &   0.21 &\nodata &\nodata &  11.99 &  27.42 &  22.95 &    8.67 &   9.00 \\
NGC\,5860     &   0.69 &   1.58 &   0.96 &\nodata &\nodata &   6.81 &    4.50 &   4.95 \\
NGC\,6090     &\nodata &  11.70 &   8.50 &  77.98 &  92.32 & 107.20 &   42.67 &  45.25 \\
NGC\,6217     &   0.17 &   1.20 &   1.08 &   3.20 &   3.83 &   7.17 &    3.07 &   3.42 \\
NGC\,6643     &   0.10 &   2.06 &   1.96 &   0.50 &   7.05 &   9.88 &    4.23 &   4.40 \\
UGC\,11284    &   2.40 &\nodata &\nodata &  73.28 &\nodata & 125.50 &   59.56 &  60.68 \\
NGC\,6753     &\nodata &   4.10 &   4.10 &\nodata &  18.74 &\nodata &    9.35 &  10.60 \\
TOL\,1924-416 &1.17$^{a}$&   3.12 &   3.08 &   3.83 &\nodata &\nodata &    0.82 &   1.32 \\
NGC\,6810     &\nodata &   1.23 &   0.79 &\nodata &  14.70 &\nodata &    0.61 &   0.90 \\
ESO\,400-G\,43&   3.29 &   9.19 &   8.62 &  11.79 &  10.94 &  18.18 &    4.84 &   7.19 \\
NGC\,7496     &\nodata &   1.29 &   0.81 &   1.35 &\nodata &   2.20 &    1.73 &   1.97 \\
NGC\,7552     &\nodata &   2.64 &   1.55 &  10.79 &  16.98 &  12.80 &   12.86 &  13.28 \\
MRK\,323      &   0.49 &   1.06 &   1.06 &   6.29 &\nodata &  13.76 &    6.52 &   6.77 \\
NGC\,7673     &\nodata &   1.31 &   1.31 &   9.67 &  12.36 &  11.33 &    4.88 &   5.58 \\
NGC\,7714     &\nodata &   4.03 &   4.03 &  12.32 &  18.92 &  15.80 &    7.17 &   7.85 \\
MRK\,332      &   0.25 &   1.63 &   0.88 &   2.98 &\nodata &   6.83 &    2.81 &   3.12 \\
\enddata
\tablenotetext{.}{column 1: galaxy name; column 2: UV star formation rate;
columns 3 and 4: star formation rates obtained from H$\alpha$ and H$\alpha$
corrected for [NII] contamination; columns 5, 6 and 7: star formation rates
from radio 8.4~GHz, 4.85~Ghz and 1.4~GHz; column 8: F$_{IR}$ (8$\mu$m to 1mm)
star formation rates, calculated using the F$_{IR}$ fluxes obtained using IRAS
and ISO measurements, whenever available, or extrapolated from IRAS
measurements only (Sanders \& Mirabel 1996); column 9: star formation rates
based on the Starburst Bolometric flux, calculated adding F$_{IR}$,
ultraviolet and B band fluxes. Both UV and H$\alpha$ fluxes have been corrected
for foreground Galactic extinction, as listed in Table 1 of paper 1.}
\tablenotetext{a}{UV measurements from IUE}
\label{tabint}
\end{deluxetable}

\begin{deluxetable}{lrrrrr}
\tabletypesize{\scriptsize}
\tablewidth{0pc}
\tablecaption{Matched Aperture Star Formation Rates}
\tablehead{
\colhead{Name}&
\colhead{UV}&
\colhead{H$\alpha$}&
\colhead{H$\alpha^{cor}$}&
\colhead{8.4~GHz}&
\colhead{1.4~GHz}\\
\cline{2-6}\\
\colhead{}&
\multicolumn{5}{c}{(M$_{\sun}$ yr$^{-1}$)}\\
\colhead{(1)}&
\colhead{(2)}&
\colhead{(3)}&
\colhead{(4)}&
\colhead{(5)}&
\colhead{(6)}}
\startdata
MRK\,555      &    0.59 &   1.12 &   0.82 &   4.04 &    11.26  \\
NGC\,1569     &    0.03 &   0.11 &   0.10 &   0.02 &     0.10  \\
NGC\,1667     &    0.69 &   1.74 &   0.94 &  11.24 &    17.98  \\
NGC\,1672     &    0.10 &   0.35 &   0.35 &\nodata &    14.69  \\
NGC\,1741     &    2.08 &   3.08 &   2.72 &   5.90 &    11.81  \\
NGC\,3079     &    0.02 &\nodata &\nodata &  19.39 &    45.54  \\
NGC\,4088     &    0.02 &   0.12 &   0.09 &   0.48 &     1.37  \\
NGC\,4214     &    0.03 &   0.09 &   0.09 &   0.05 &     0.03  \\
NGC\,4861     &    0.53 &   0.60 &   0.60 &   0.85 &     0.71  \\
NGC\,5383     &    0.19 &   3.27 &   3.27 &   3.27 &     6.24  \\
MRK\,799      &    0.12 &   0.72 &   0.48 &   6.84 &    11.73  \\
NGC\,5669     &    0.08 &\nodata &\nodata &\nodata &     1.83  \\
NGC\,5676     &    0.06 &\nodata &\nodata &   2.32 &     6.56  \\
NGC\,5713     &    0.21 &\nodata &\nodata &   7.52 &    14.40  \\
NGC\,5860     &    0.69 &   1.58 &   0.96 &\nodata &     6.81  \\
NGC\,6217     &    0.17 &   0.47 &   0.42 &   3.17 &     7.11  \\
NGC\,6643     &    0.09 &   0.30 &   0.29 &   0.11 &     2.20  \\
UGC\,11284    &    2.40 &\nodata &\nodata &  31.00 &    53.11  \\
ESO\,400-G\,43&    3.29 &   9.10 &   8.53 &  11.25 &    17.34  \\
MRK\,323      &    0.49 &   0.78 &   0.78 &   4.82 &    10.56  \\
MRK\,332      &    0.25 &   1.11 &   0.60 &   1.99 &     4.55  \\
IC\,1623      &    5.37 &\nodata &\nodata & 135.30 &   249.00  \\
\enddata
\tablenotetext{.}{Column 1: galaxy name; column 2: UV star formation rate;
columns 3 and 4: H$\alpha$ and H$\alpha$ corrected for [NII] contamination
star formation rates; columns 5 and 6: star formation rates from radio
8.4~GHz and 1.4~GHz.}
\label{tabmatch}
\end{deluxetable}

\clearpage
 
\begin{figure}
\plotone{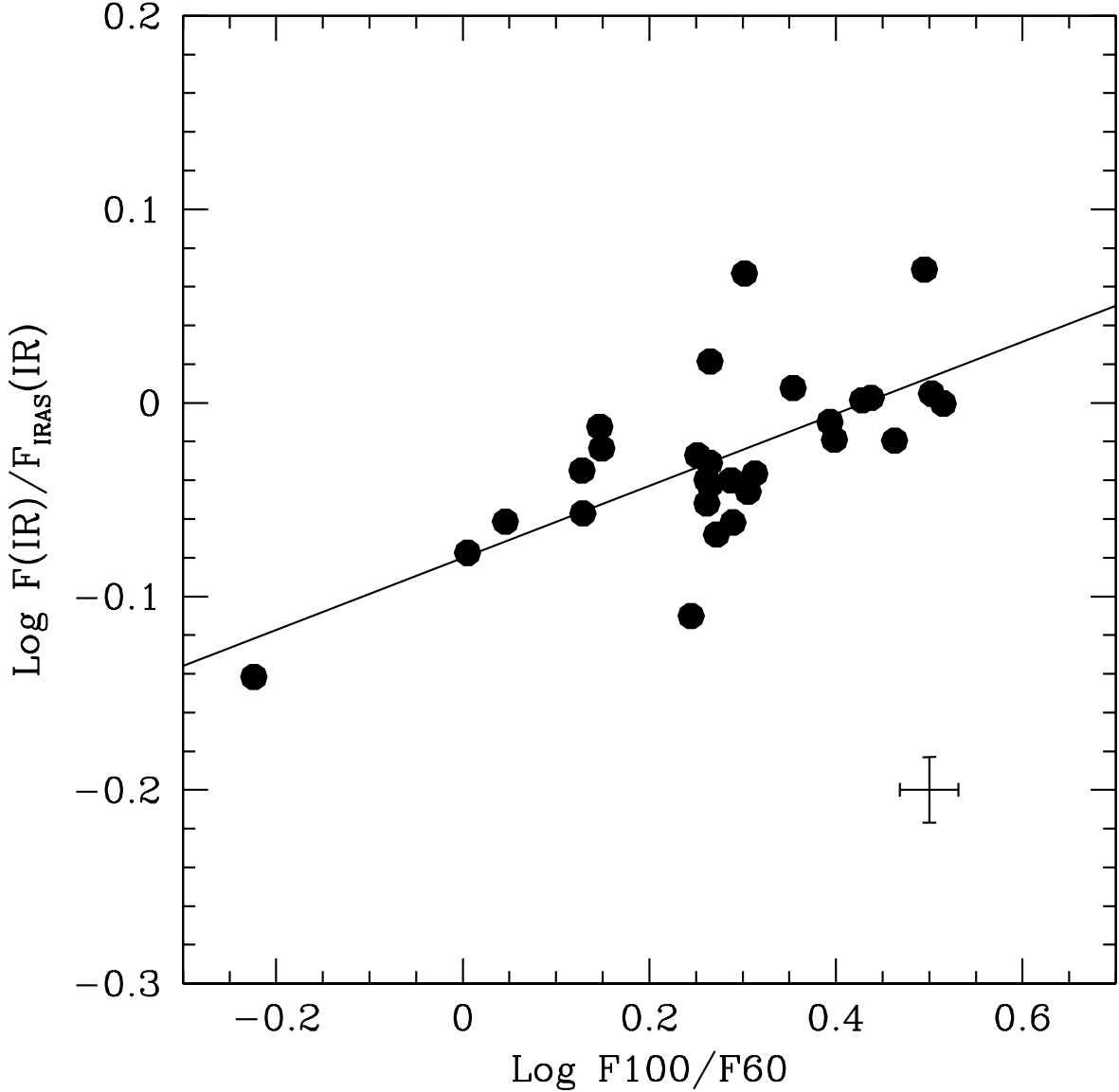}
\caption{The logarithm of the ratio between the infrared flux (8$\mu$m--1mm)
measured using IRAS and ISO data to the infrared flux extrapolated using
only IRAS data, as a function of the logarithm of the ratio between the
IRAS 100$\mu$m and 60$\mu$m fluxes. The solid line represents the linear
regression fit to the data points, which has a Spearman $\rho=0.656$, 
corresponding to a probability of 0.05\% that a correlation is not present.
The median errorbar is shown in the bottom right corner of the figure.}
\label{fig2}
\end{figure}

\begin{figure}
\plotone{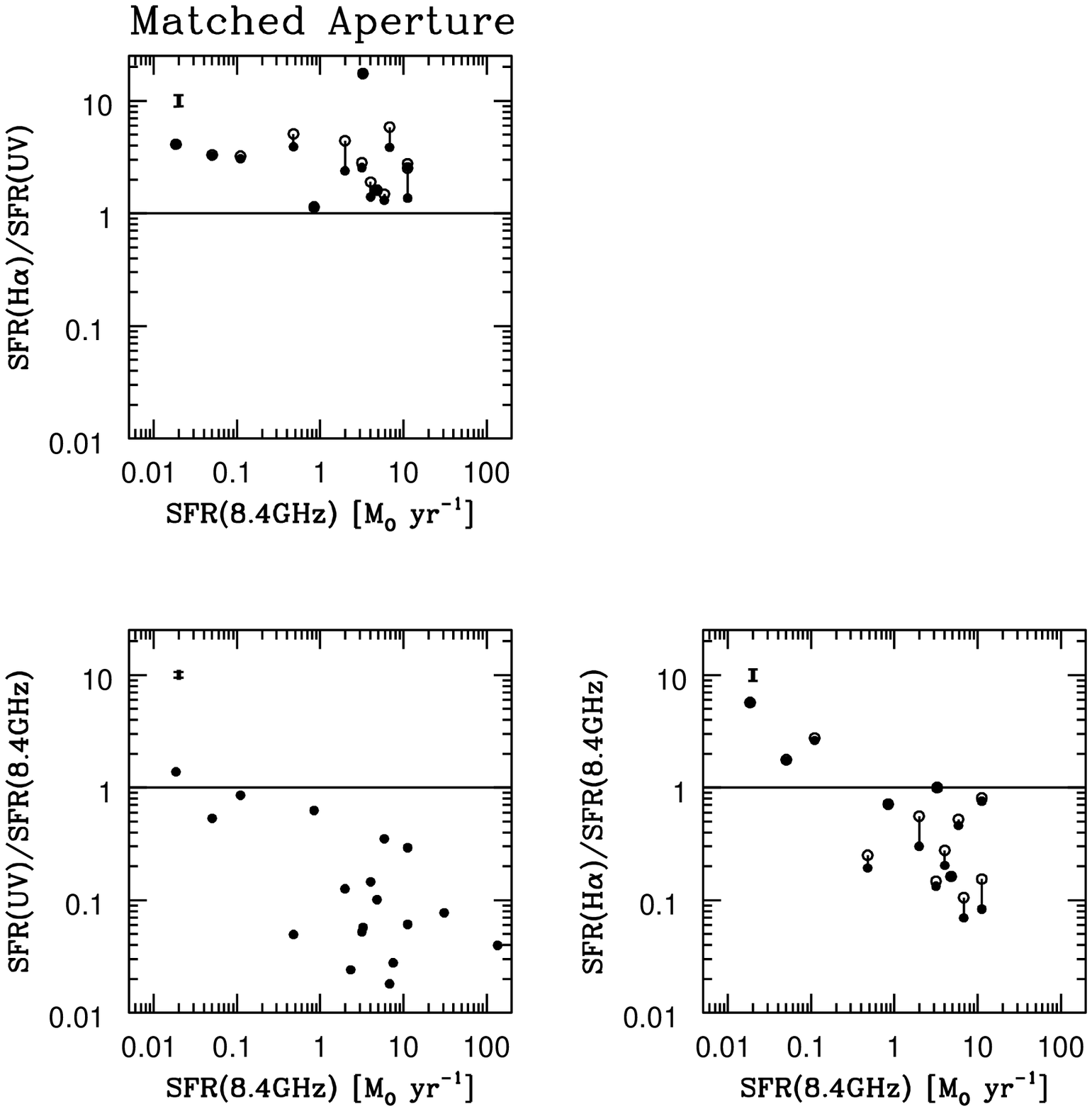}
\caption{SFR(H$\alpha$)/SFR(UV) (top), SFR(UV)/SFR(8.4~GHz) (bottom left),
and SFR(H$\alpha$)/SFR(8.4~GHz) (bottom right), as a function of SFR(8.4~GHz),
for observed fluxes measured in STIS-matched apertures. SFR(8.4~GHz)
was calculated using equation (7). In the top and bottom right panels
the open and filled circles represent the H$\alpha$
measurements uncorrected and corrected for [NII] contamination, respectively.
Notice that in some cases the open dots are not seen, because of small [NII]
correction factors. A one to one correlation between each pair of
quantities is shown by the horizontal line in these plots. Median errorbars
are shown in the top left corner of each panel.}
\label{fig3}
\end{figure}

\begin{figure}
\plotone{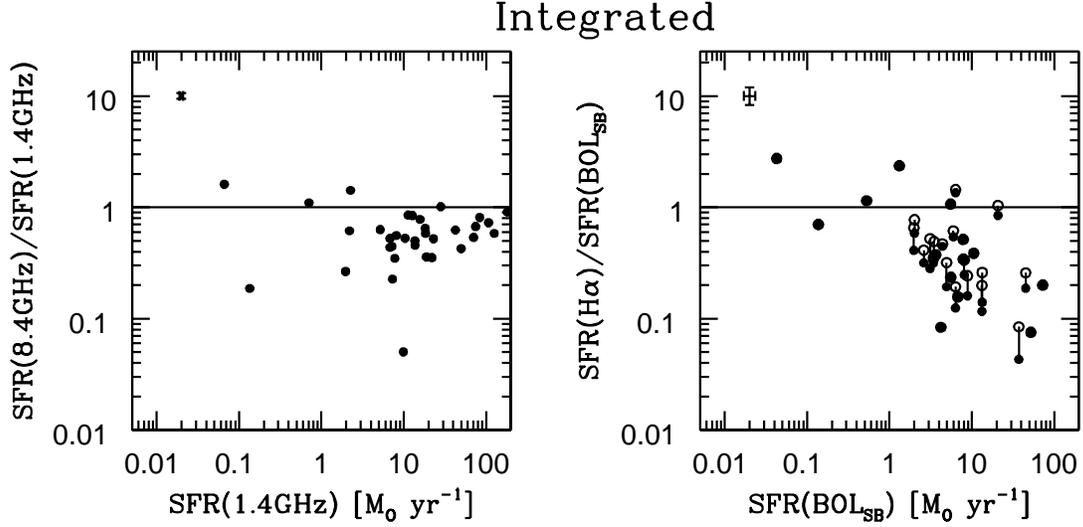}
\caption{Comparison between different star formation rates calculated using the
integrated light of the galaxies. The left panel shows the distribution of
SFR(8.4~GHz)/SFR(1.4~GHz) as a function of SFR(1.4~GHz), while the right
one shows SFR(H$\alpha$)/SFR(BOL$_{SB}$), as a function of SFR(BOL$_{SB}$)
The two symbols in the right panel indicate H$\alpha$ measurements before (open
circles) and after (filled circles) the correction for [NII] contamination.
A one to one correlation between the quantities presented in these plots
is shown by the horizontal line. Median errorbars
are shown in the top left corner of each panel.}
\label{fig4}
\end{figure}

\begin{figure}
\plotone{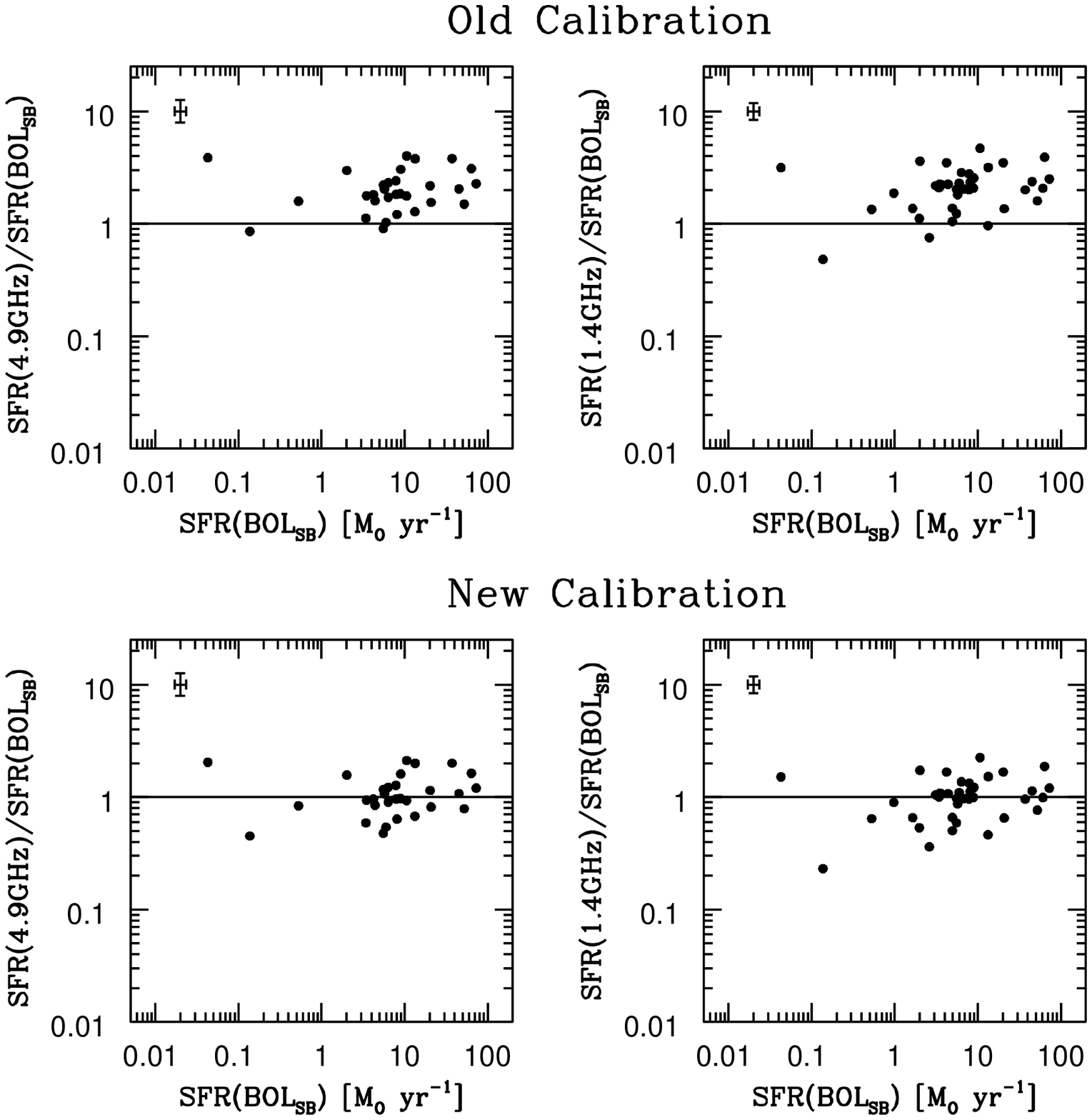}
\caption{Comparison of SFR(4.9~GHz)/SFR(BOL$_{SB}$) and
SFR(1.4~GHz)/SFR(BOL$_{SB}$) as a function of SFR(BOL$_{SB}$), left and right
panels respectively. The top panels show the values calculated using the old
radio SFR calibration (Equation 7) and the bottom ones show the values
calculated using our new calibration (Equations 
8 and 10). Median errorbars are shown in the top left corner of each panel
and the one to one correlation is shown as a horizonthal line.}
\label{fig5}
\end{figure}

\begin{figure}
\plotone{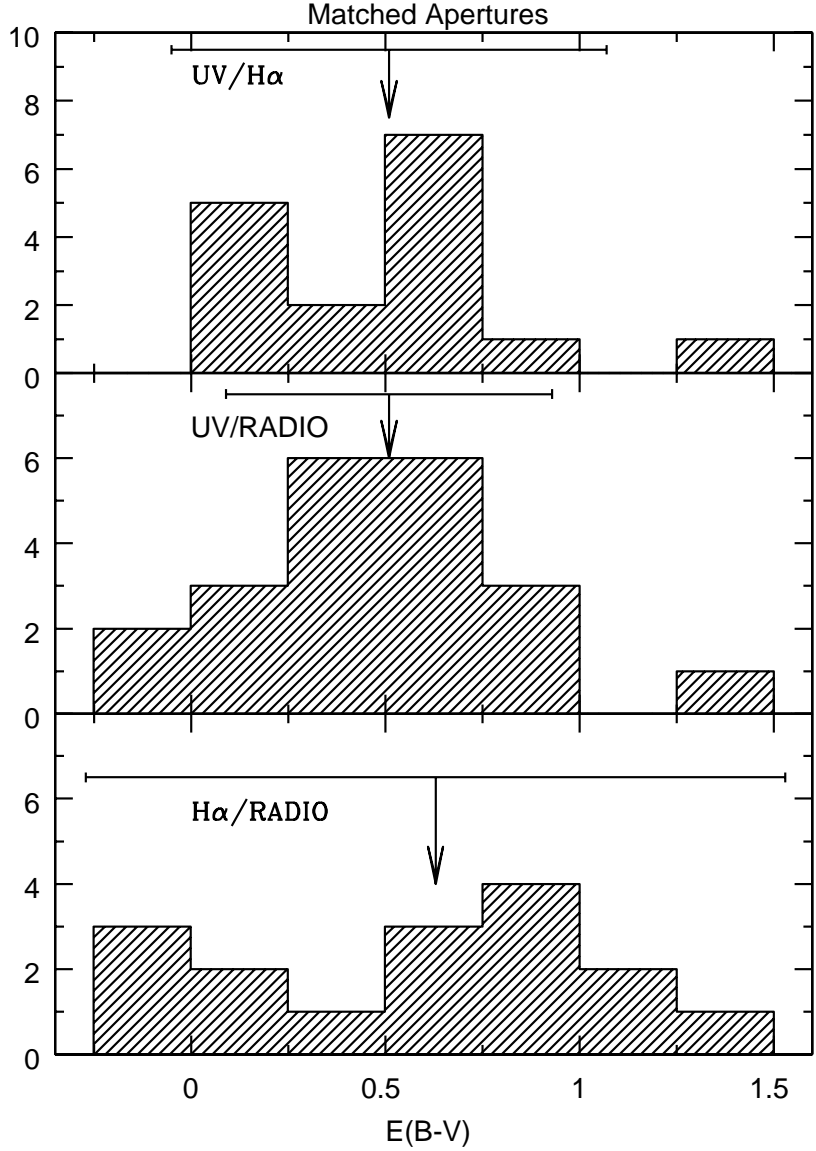}
\caption{Histograms of the color excess E(B-V) calculated using the
discrepancy between each pair of UV, H$\alpha$ and radio 8.4~GHz SFR, measured
in matched apertures (see Figure~3). Here we use the new SFR(8.4~GHz)
calibration discussed in Section~4.3. The arrow indicates the position
of the median and the error bar shows the interquartile range. The E(B-V)
values were calculated assuming a starburst extinction curve for the UV
(Calzetti et al. 2000) and the Galactic one for H$\alpha$ (Cardelli, Clayton \&
Mathis 1989). The median UV over radio ratio corresponds to A$_{UV}\sim2.3$
and the median H$\alpha$/radio corresponds to A$_{H\alpha}\sim1.5$. Although
E(B-V)$_{UV}$ is smaller than E(B-V)$_{H\alpha}$, A$_{UV}$ is still larger
than A$_{H\alpha}$, thus explaining the trend in Figure~3.}
\label{fig6}
\end{figure}

\begin{figure}
\plottwo{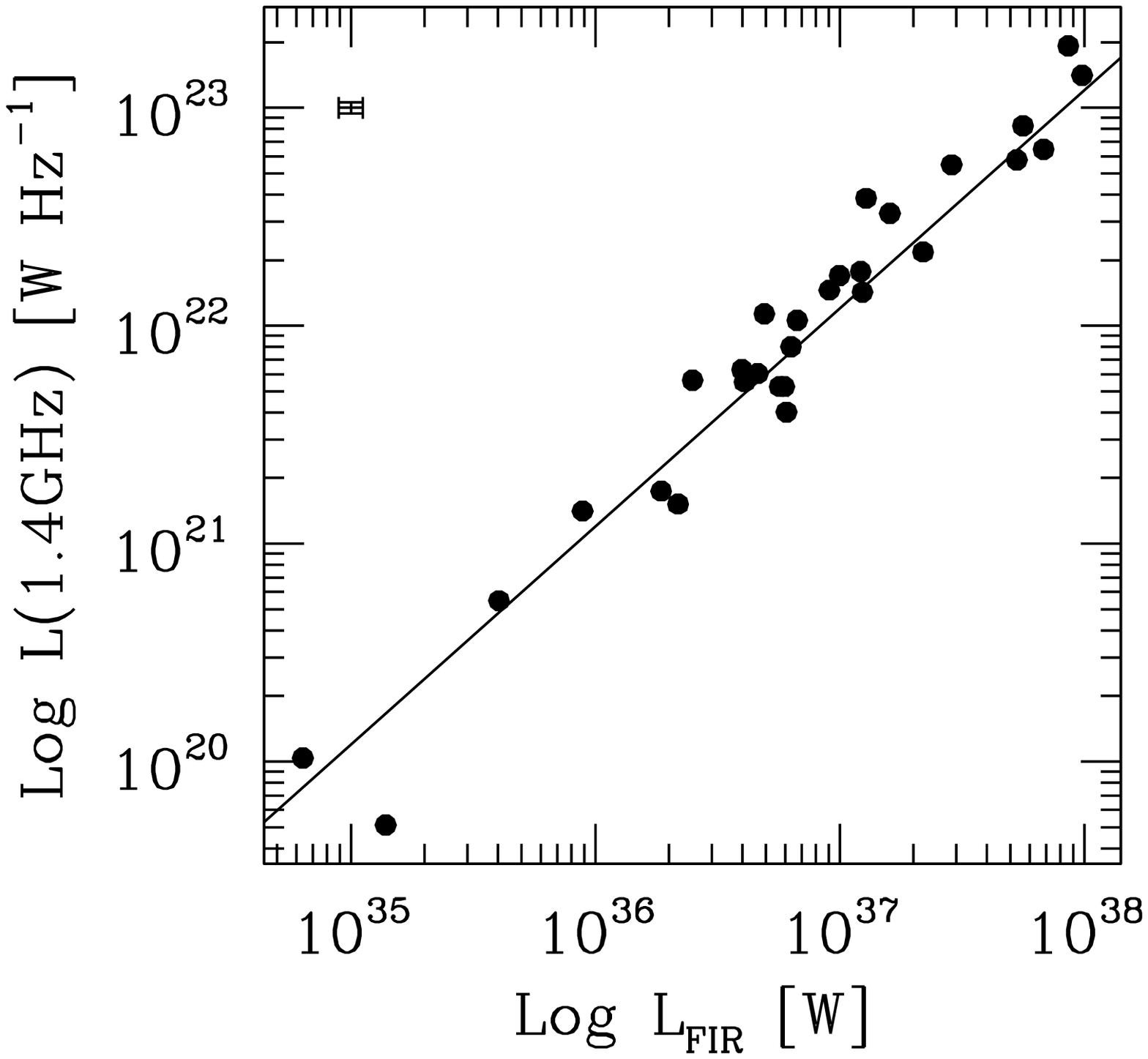}{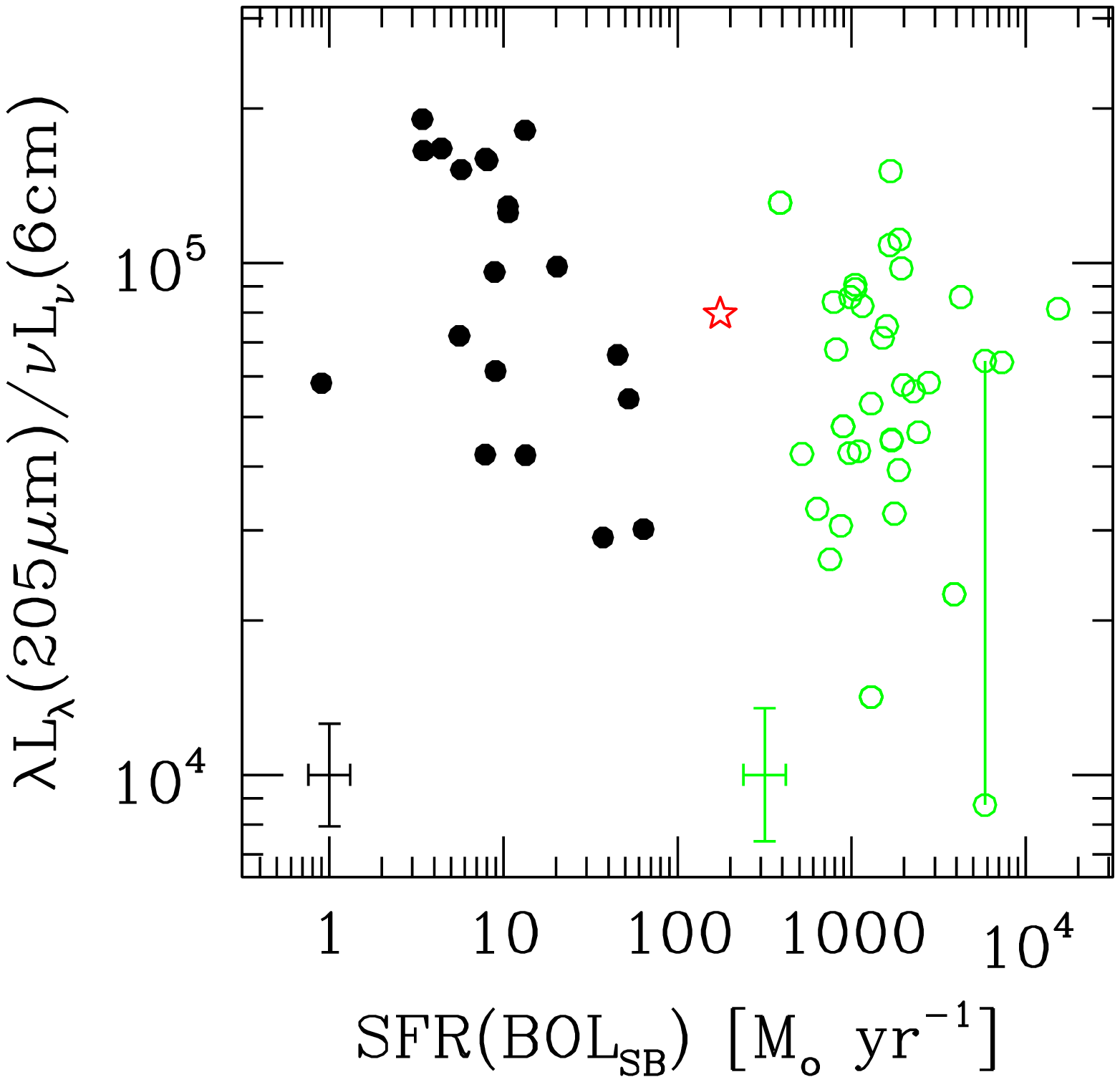}
\caption{The left panel shows the radio 1.4~GHz$\times$FIR diagram for our
galaxies, with the relation from Condon et al. (1991) shown as a solid line
(median errorbar is shown in the top left corner). The right panel shows the
infrared L$_{\lambda}$(205$\mu$m) over radio L$_{\nu}$(6~cm) flux density
ratio as a function of star formation rate. The filled dots represent
our data points, the red star represents Arp\,220 and the open green circles
represent SCUBA sources. One of the SCUBA sources (SMM\,J\,163650.0$+$405733)
has two radio fluxes related to it (Ivison et al. 2002), as described in the
text, so we show the two values connected by a vertical bar. The black
errorbar (bottom left) shows the median error for our sample, and the green
one (bottom middle) corresponds to the median error of the SCUBA sources.}
\label{fig7}
\end{figure}

\begin{figure}
\plotone{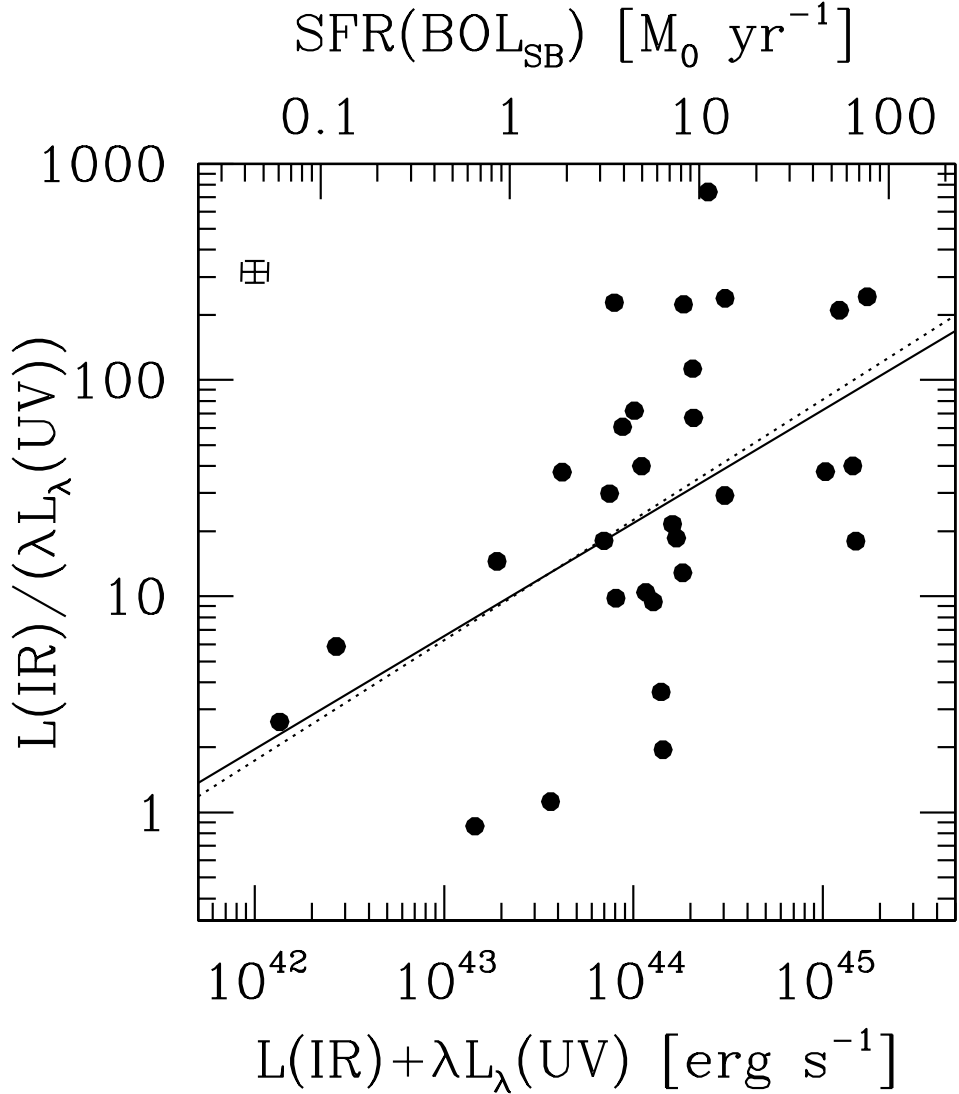}
\caption{The IR/UV luminosity ratio as a function of the sum of the
two quantities. We make the approximation
BOL$_{SB}\sim$L$_{IR}+\lambda$L(UV), in order to convert the X-axis to
the SFR(BOL$_{SB}$) value given on the top axis. Those galaxies
without ISO measurements were corrected using the equation from the
previous figure. UV data for 9 of the galaxies in this plot were
obtained from IUE. The solid line represents the best fit relation of Heckman
et al. (1998), while the dotted one represents the best fit to our data.
The median errorbar is shown in the top left corner.}
\label{fig8}
\end{figure}

\begin{figure}
\plottwo{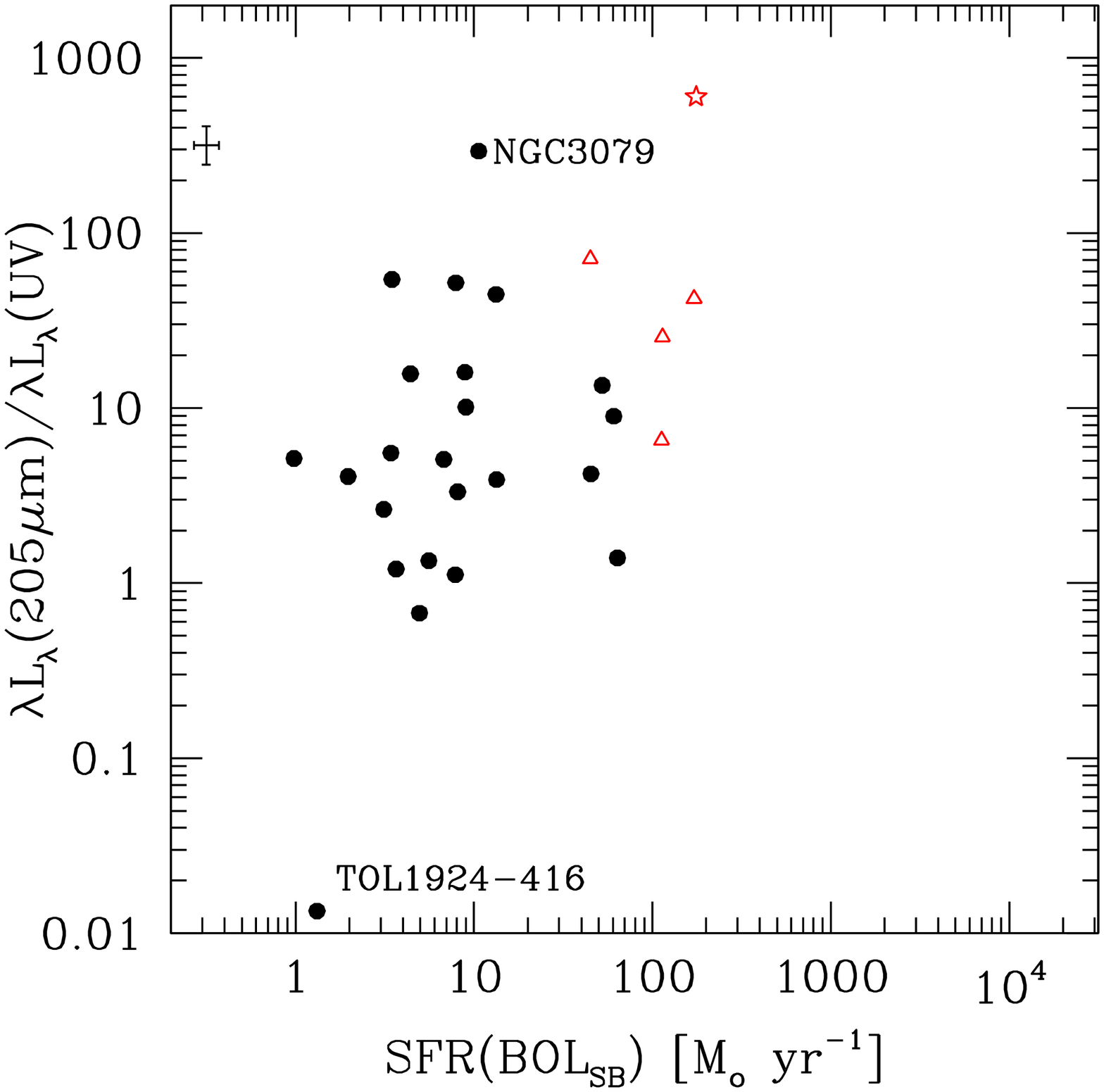}{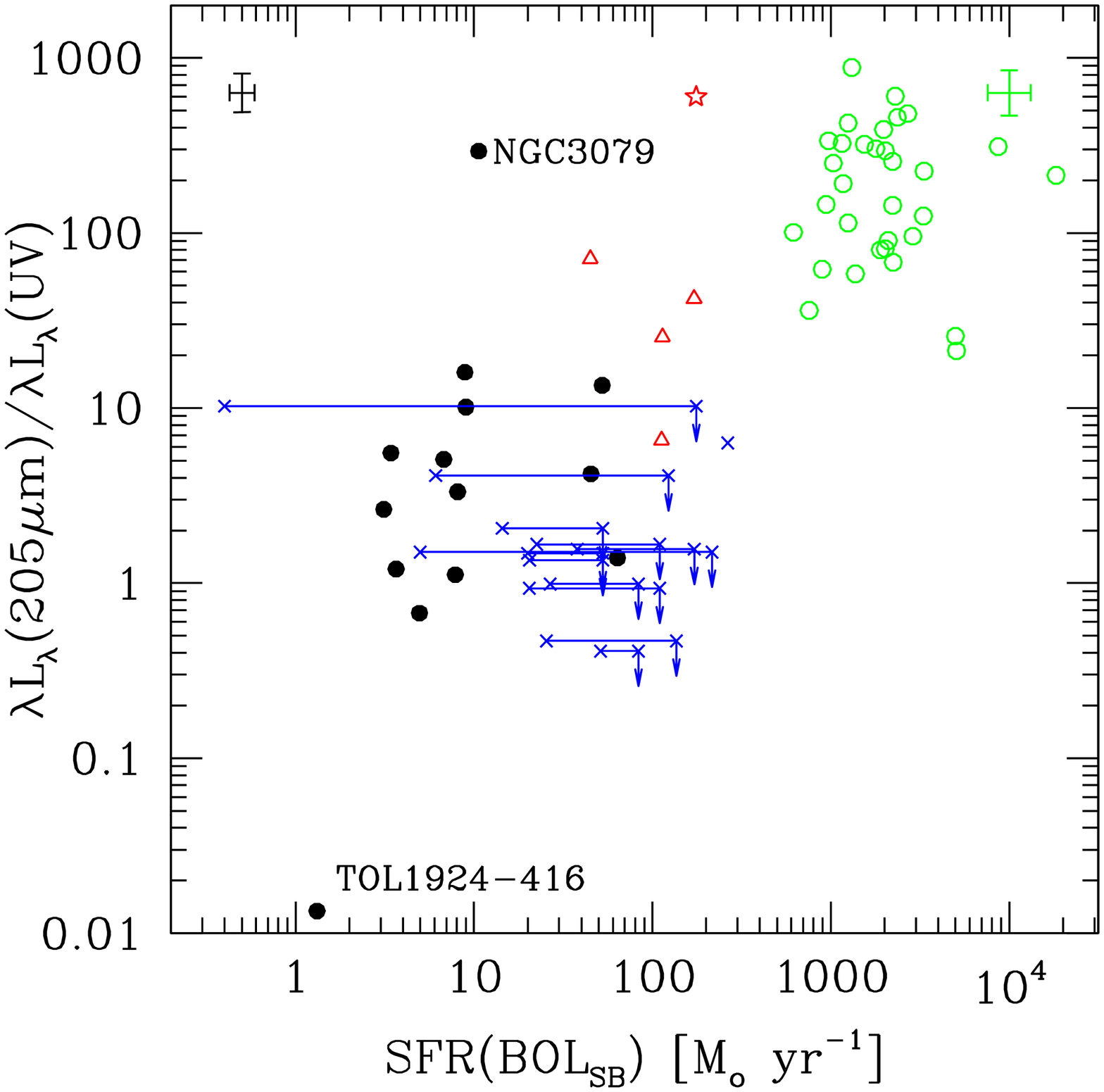}
\caption{The monochromatic IR flux to UV ratio as a function of the bolometric
star formation rate for all galaxies in our sample (left) and only those
galaxies for which more than 50\% of the radio emission is contained within the
STIS aperture (right). Filled dots represent our data points, the red star is
Arp\,220, the red triangles are ULIGs (Goldader et al. 2002; Klaas et al. 2001;
Becker et al. 1991), green open circles represent SCUBA sources (Chapman et al.
2003, 2005) and crosses represent LBGs (Chapman et al. 2000). The black point
with small $\lambda$L$_{\lambda}$(205$\mu$m)/$\lambda$L$_{\lambda}$(UV)
corresponds to TOL\,1924--416, which is
a dwarf galaxy with a young starburst and almost no emission from cold dust,
while the point with the largest ratio is NGC\,3079, a high inclination galaxy
known to harbor an AGN. In the case of LBGs, with the exception of one source,
they are all upper limits on the Y axis. Along the X axis the LBG's have
SFR's that are bracketed by the upper limits from SFR(210$\mu$m) and the
lower limits given by the observed UV. The median errorbar of our sample is
shown in black in the top left corner, while in the case of SCUBA sources
it is shown in green in the top right corner.}
\label{fig9}
\end{figure}

\end{document}